\documentclass[showpacs,amsmath,amssymb,aps,twocolumn,superscriptaddress,prx]{revtex4-2}
\usepackage{mathtools}
\usepackage{algorithm}
\usepackage{algpseudocode}
\usepackage[pdftex]{graphicx} \graphicspath{{}}% Include figure files
\usepackage{grffile} % change algorithm for searching thru graphics extensions
\usepackage{comment}
\usepackage{braket}
\usepackage{rotating}

\usepackage{float}
\usepackage{dcolumn}% Align table columns on decimal point
\usepackage{bm}% bold math
\usepackage[pdftex,colorlinks=true]{hyperref}
\usepackage{soul}
\hypersetup{
	colorlinks=true,
	linkcolor=blue,
	urlcolor=cyan,
}

\usepackage{color}
\usepackage{xcolor}
\definecolor{ForestGreen}{RGB}{34, 139, 34}

\newcommand{\AI}[1]{\ifmmode\textsf{\color{blue}\textnormal{$#1$}}\else\textsf{\color{blue}#1}\fi}

\usepackage[normalem]{ulem}

\newcommand{\affA}{State Key Laboratory of Artificial Microstructure and Mesoscopic Physics, School of Physics, Peking University, Beijing 100871, China}
\newcommand{\affC}{Frontiers Science Center for Nano-Optoelectronics, Peking University, Beijing 100871, China}
\newcommand{\affD}{Collaborative Innovation Center of Extreme Optics, Shanxi University, Taiyuan 030006, China}

\begin{document}

\title{{Resonance and Transition-Channel Competition in Cavity-Assisted Floquet-State Stabilization}}

\author{Zhuocheng Ma}
 \affiliation{\affA}
 \author{Hongzheng Zhao}
	\email{hzhao@pku.edu.cn}
 \affiliation{\affA}
 \author{Liang-You Peng}
 \email{liangyou.peng@pku.edu.cn}
 \affiliation{\affA }
 \affiliation{ \affC }
 \affiliation{ \affD}
	\date{\today}

\begin{abstract}

Static coupling to a dissipative cavity has been proposed as a route to autonomously stabilize Floquet eigenstates. The cavity provides an energy-selective relaxation channel: when its photon energy matches a quasienergy difference modulo drive quanta, it mediates transitions between Floquet states, followed by irreversible photon loss. Here, we show that at intermediate driving frequencies, this resonance condition alone is not sufficient for stabilization. Instead, the steady state is controlled by the interplay between resonance selection and the cavity-induced transition strengths between Floquet states, which can either cooperate or compete. This interplay leads to successful stabilization, mixed steady states despite resonance, and complete breakdown near Floquet-zone hybridization. We further show that the same mechanism extends to multilevel spin-\(S\) systems, providing a general framework for understanding when static dissipation can reliably stabilize driven quantum states.

\end{abstract}
	\maketitle
\let\oldaddcontentsline\addcontentsline
\renewcommand{\addcontentsline}[3]{}

% 这个tex要用pdflatex编译

\section{Introduction}

Time-dependent driving protocols have emerged as powerful tools for engineering novel non-equilibrium phenomena, including discrete time crystals and quasicrystals~\cite{PhysRevB.100.134302, PhysRevB.99.220303, PhysRevLett.123.150601, e27060609, feng2025tunablediscretequasitimecrystal}, topological energy pumping~\cite{PhysRevX.7.041008, PhysRevLett.127.166804, PhysRevB.99.064306}, and dynamical localization in the frequency domain~\cite{PhysRevB.105.L220202, huWannierStarkLocalization2025, PhysRevE.93.062205, PhysRevB.109.L161104}. These protocols have been experimentally realized on platforms including nitrogen-vacancy centers~\cite{PhysRevX.15.011055} and superconducting quantum processors~\cite{zhangDigitalQuantumSimulation2022a, liuPrethermalizationRandomMultipolar2026a}.

A central ingredient in exploiting such driven systems is the controlled preparation of their Floquet eigenstates. In the high-frequency regime, the stroboscopic dynamics can be described by an engineered effective Hamiltonian, enabling synthetic gauge fields and  tunable interactions~\cite{Goldman2014, Eckardt_2015, Aidelsburger2013, Kyriienko2018, Nguyen2024, Liang2024}; preparing its desired eigenstates then provides access to the corresponding engineered many-body physics~\cite{Bukov04032015, Oka2019}. Beyond this regime, controlled occupation of Floquet eigenstates is likewise essential for realizing intrinsically dynamical phases with no static counterpart, including Floquet topological insulators and non-Abelian Floquet spin liquids~\cite{Jotzu2014, Kitagawa2010, PhysRevLett.110.200403, Rudner2020, Harper2020, Kalinowski2023, Sun2023}.

{In practice, however, the system is typically initialized in a simply preparable state, such as a product state that overlaps with many Floquet eigenstates. Selectively populating a target Floquet eigenstate while suppressing occupation of the others therefore remains a challenging task. Even once such a state is prepared, maintaining it presents a second fundamental challenge: coupling to the environment, through spontaneous emission, dephasing, and other noise channels, induces incoherent transitions between Floquet eigenstates and generally drives the system away from a pure Floquet state. Consequently, the steady state of the driven open system is generically mixed rather than a single Floquet eigenstate~\cite{Shirai2015}. }

% \hz{Throughout this work, we refer to this environment-induced loss of Floquet-state purity as \emph{heating}, to distinguish it from the intrinsic heating toward infinite temperature that occurs in generic closed ergodic Floquet systems~\cite{Lazarides2014,DAlessio2014}. We use \emph{stabilization} to mean the dynamical purification of the system into a Floquet eigenstate.}

Considerable effort has been devoted to this problem. {One potential approach} is to couple the driven system to a carefully designed dissipative environment so that excess energy absorbed from the drive is removed and the system relaxes toward a desired non-equilibrium steady state. Two classes of strategies have been explored: engineering the properties of a thermal bath to dissipatively prepare target Floquet states~\cite{schnellDissipativePreparationFloquet2024, wanckelDissipativeFloquetEngineering2026, steinfadtDissipationAssistedPreparationFloquetLaughlin2026}, and coupling the driven system to a lossy auxiliary cavity whose dissipation autonomously stabilizes a Floquet eigenstate~\cite{PhysRevLett.129.233601, ritterAutonomousStabilizationFloquet2025, ramanDissipativeStabilizationFloquetEngineered2026}. In this work, we focus on the latter approach. In particular, it was demonstrated that static coupling to an unmodulated but lossy auxiliary system can stabilize a driven system in a target Floquet eigenstate~\cite{ritterAutonomousStabilizationFloquet2025}. 
The mechanism
is illustrated in Fig.~\ref{process}(b): When the cavity frequency satisfies the resonance condition, the cavity photon energy matches the quasienergy splitting modulo integer multiples of the driving frequency. The spin--cavity interaction then resonantly couples $|u_+,0\rangle$ to $|u_-,1\rangle$ (blue arrow), where $|u_\pm,n\rangle$ denotes a Floquet state $|u_\pm\rangle$ accompanied by $n$ cavity photons. Subsequent photon dissipation then transfers $|u_-,1\rangle$ to $|u_-,0\rangle$, thereby stabilizing the spin in the target Floquet state $|u_-\rangle$. We refer to this transition-dissipation pathway as the \emph{Blue process}. It was further argued that this static-dissipation strategy can stabilize Floquet quasienergy states over a wide range of driving frequency.

The Blue process, however, need not be the only transition-dissipation pathway available. Other channels can coexist with it and favor different Floquet eigenstates. For instance, as shown in Fig.~\ref{process}(b), a competing \emph{Red process} transfers population in the opposite direction and tends to stabilize the other Floquet eigenstate, with the drive supplying the required energy. This competing process is strongly suppressed in both the high-frequency and adiabatic limits, where the Blue process therefore dominates. At intermediate driving frequencies, however, the Red process can become appreciable and compete with the Blue process, preventing purification into a single Floquet eigenstate. This problem can become even more severe in multilevel or many-body Floquet systems, where both the quasienergy spectrum and the available transition pathways become considerably more complex. This raises the central question of this work: is satisfying the resonance condition alone sufficient for stabilization, or must additional conditions be taken into account?

In this work, we address this question by revealing a cooperation--competition mechanism that governs cavity-assisted Floquet stabilization. At intermediate driving frequencies
% \hz{you also mention in low-frequency regime the following discussion still applies; however, in last paragraph, you also say Red process will be strongly suppressed in adiabatic limit. Do you mean different things when you use "adiabatic limit" and "low-frequency regime"?}\zc{OK, let's use intermediate driving
% frequencies }, 
the resonance condition alone no longer determines the dominant transition process, and the transition matrix elements become essential in determining the relative strengths of the Red and Blue channels. Their interplay gives rise to three distinct regimes. In the \emph{cooperative} regime, the dominant transition channel is also resonantly enhanced, and stabilization succeeds. In the \emph{competitive} regime, the resonance condition favors one channel while the transition matrix elements favor the other, so stabilization fails even though the resonance condition is satisfied. Finally, in the \emph{near-degenerate} regime, the competing channels become comparable and stabilization breaks down for all cavity frequencies.

We demonstrate this mechanism first in a periodically driven two-level spin, which provides the minimal setting in which the competition between transition channels can be clearly resolved. We then generalize the framework to multilevel systems, where multiple Floquet levels and transition pathways participate. Focusing on a single spin-\(S\) model~\cite{PhysRevA.108.063716}, we show that its \(\mathfrak{su}(2)\) structure strongly constrains both the quasienergy spectrum and the allowed transitions between Floquet states. The uniformly spaced unfolded Floquet spectrum allows a single cavity frequency to resonantly address all neighboring levels, while the transition matrix elements generically favor the same transition direction between each neighboring pair. The cooperation or competition between these two ingredients produces the same three regimes found in the two-level system.

The remainder of the paper is organized as follows. In Sec.~II, we introduce the two-level model and develop the cooperation--competition mechanism underlying stabilization and its breakdown. In Sec.~III, we generalize this framework to multilevel spin-\(S\) systems. Section~IV concludes with a discussion of our results.

\section{Two-level Floquet system coupled to a dissipative cavity}

We begin with a two-level Floquet system---a periodically driven spin coupled to a dissipative cavity. We analyze two competing transition-dissipation pathways and show how their interplay determines whether the spin system relaxes to a nearly pure Floquet eigenstate or to a high-entropy mixed state.

\subsection{Quasienergies and quasienergy states}
\label{sec.Floquet}

% We first introduce the quasienergy states and quasienergies of a Floquet system~\cite{7l91-gw77}, 
We consider a periodically driven Hamiltonian $H_S(t)=H_S(t+T)$ with driving period $T$. The quasienergy states are special solutions to the Schr\"odinger equation and take the form
\begin{equation}
    |\psi_\pm(t)\rangle=e^{-i\varepsilon_\pm t}|u_\pm(t)\rangle,
\end{equation}
where $\varepsilon_\pm$ are the quasienergies and $|u_\pm(t)\rangle$ are periodic Floquet modes satisfying $|u_\pm(t+T)\rangle=|u_\pm(t)\rangle$. Substituting into the Schr\"odinger equation yields
\begin{equation}
    [H_S(t)-i\partial_t ]|u_\pm(t)\rangle = \varepsilon_\pm |u_\pm(t)\rangle.
\end{equation}
Quasienergies are defined modulo the driving frequency $\omega=2\pi/T$; specifically, the transformation
\begin{equation}
    \begin{matrix}
 \varepsilon_\pm \to  \varepsilon_\pm+n\omega=\varepsilon^{(n)}_\pm,\\
|u_\pm(t)\rangle \to e^{in\omega t}|u_\pm(t)\rangle =|u^{(n)}_\pm(t)\rangle,
\end{matrix}
\end{equation}
yields identical wavefunction $|\psi_\pm(t)\rangle$
% $|\psi_\pm(t)\rangle = e^{-i(\varepsilon_\pm +n\omega) t} |u^{(n)}_\pm(t)\rangle$ 
for any integer $n$. 
% The Floquet modes are also eigenstates of the one-period evolution operator,
% $$
% U(T)=\mathcal{T}e^{-i\int^{T+t_0}_{t_0} dt H_S(t)},
% $$
% with eigenvalues $e^{-i\varepsilon_\pm T}$, where $\mathcal{T}$ denotes time ordering. 
For convenience, we choose the quasienergies to lie in the first Floquet zone, $|\varepsilon_\pm|<\omega/2$, in analogy to the first Brillouin zone of quasimomentum, and we denote the quasienergy splitting by $\delta\varepsilon\equiv\varepsilon_+-\varepsilon_-$.

\subsection{The model}

\label{sec.model}

We consider a two-level spin driven by a periodic magnetic field and coupled to a dissipative cavity. The Hamiltonian involves three parts,
\begin{align}
 &H_{S}(t)=-A \cos(\omega t)\sigma_{y}+\frac{\Omega}{2} \sigma_{z} ,
 H_{C}=\Delta a^\dagger a ,\\
&H_{I}=g(a^\dagger \sigma^-  + a\sigma^+ ) ,
\label{Fmodel}
\end{align}
where $H_S(t)$ is the two-level system Hamiltonian with driving frequency $\omega$ and amplitude $A$; $\Omega$ is a constant; $\sigma_i$ ($i\in\{x,y,z\}$) are Pauli matrices, with $\sigma^\pm=(\sigma_x\pm i\sigma_y)/2$ the spin ladder operators. The cavity Hamiltonian $H_C$ involves creation and annihilation operators $a^\dagger,a$, with cavity frequency $\Delta$. The interaction Hamiltonian $H_I$ describes coupling between the cavity and the spin.

In the presence of dissipation, the system's dynamics is governed by the master equation
\begin{equation}
    \partial_t \rho=-i[H_S(t)+H_C+H_I,\rho]+\kappa \mathcal{D}_{a}[\rho]+\gamma \mathcal{D}_{\sigma^-}[\rho],
    \label{rhoME}
\end{equation}
where $\kappa$ is the photon loss rate and $\gamma$ is the spin relaxation rate. Here, $\mathcal{D}_{a}[\rho]=a\rho a^\dagger -\frac{1}{2}\{ a^\dagger a,\rho\}$ describes the cavity photon dissipation, and $\mathcal{D}_{\sigma^-}[\rho]=\sigma^-\rho \sigma^+ -\frac{1}{2}\{ \sigma^+ \sigma^-,\rho\}$ describes the qubit relaxation.

To stabilize the spin into a single Floquet state of $H_S(t)$, we require the energy scale hierarchy  $|\varepsilon_+ -\varepsilon_-|,\,\Delta\gg \kappa\approx g\gg \gamma$~\footnote{Stabilization can remain effective even at exact quasienergy degeneracies ($\delta\varepsilon=0$), where this hierarchy is violated; our model provides such an example, as analyzed in Sec.~II.E.}. Since the photon number must remain a good quantum number, the interaction strength $g$ must be much smaller than both the system and cavity energy scales, so that the overall Floquet state can be approximated as a direct product $|u_\pm,n\rangle$ of the spin state and a photon number state $|n\rangle$. The cavity dissipation rate $\kappa$ must be much larger than the spin relaxation rate $\gamma$, so that the cavity can effectively remove heat generated by spontaneous emission via photon loss; however, to prevent decoupling between the system and the cavity, $\kappa$ must not significantly exceed the interaction strength~\cite{ritterAutonomousStabilizationFloquet2025}, since an excessively large $\kappa$ would suppress the very transitions it is meant to dissipate, in analogy with the quantum Zeno effect~\cite{Misra1977,Itano1990}. The quasienergy splitting $|\varepsilon_+-\varepsilon_-|$ depends on the driving frequency, so the scale of $\omega$ is indirectly constrained relative to the other parameters.

\subsection{Numerical observation}

Spontaneous emission from the spin generically drives the system to a mixed steady state, even if the spin is initialized in a Floquet eigenstate of $H_S(t)$. However, coupling to a dissipative cavity can automatically stabilize the Floquet spin system in one of the Floquet eigenstates.

\begin{figure}[!t]
    \centering
    \includegraphics[width=0.9\linewidth]{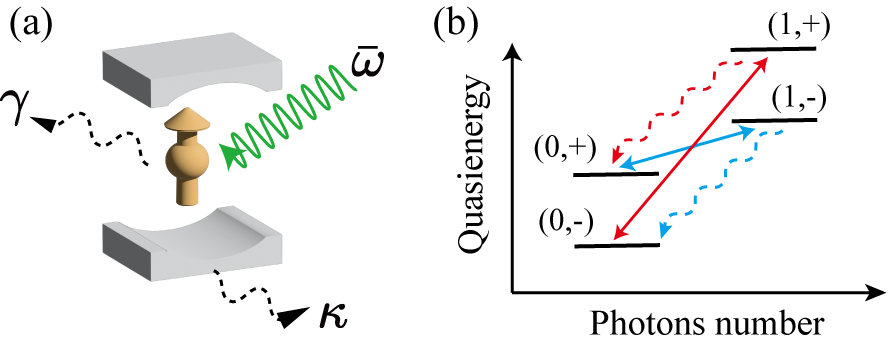}
    \caption{The dissipative cavity stabilizes the periodically driven spin. (a) Schematic of the periodically driven system. The spin is driven by an external time-dependent magnetic field that evolves periodically with frequency $\omega$; $\kappa$ denotes the cavity photon dissipation rate, which is much larger than the spin spontaneous emission rate $\gamma$.
(b) Two sets of transition-dissipation processes. Here $|u_+\rangle$ and $|u_-\rangle$ denote the Floquet excited state and Floquet ground state, respectively; the numbers $0,1$ denote photon numbers. The red lines depict the Red process: the system and cavity absorb energy from the drive to transition from $|u_-,0\rangle$ to $|u_+,1\rangle$, followed by photon dissipation to $|u_+,0\rangle$. The blue lines depict the opposite Blue process.}
    \label{process}
\end{figure}
As discussed in Ref.~\cite{ritterAutonomousStabilizationFloquet2025}, stabilization originates from the resonance between $|u_+,0\rangle$ and $|u_-,1\rangle$ followed by photon loss from $|u_-,1\rangle$ to $|u_-,0\rangle$. Concretely, when the cavity frequency $\Delta$ is comparable to the quasienergy difference between $|u_\pm\rangle$, a resonance occurs between $|u_+,0\rangle$ and $|u_-,1\rangle$, as shown by the blue solid arrow in Fig.~\ref{process}(b). The spin-cavity interaction opens a transition channel between $|u_-,1\rangle$ and $|u_+,0\rangle$; photon loss then induces the transition from $|u_-,1\rangle$ to $|u_-,0\rangle$, as illustrated by the blue wavy arrow in Fig.~\ref{process}(b). The state $|u_-,0\rangle$ thus becomes an attractor of the photonic loss, so the spin stabilizes in the Floquet ground state.
% \hz{We should call it Floquet ground state, unless you specify which BZ you are considering. }. 

In our model (Sec.~\ref{sec.model}), this stabilization mechanism indeed operates in certain parameter regimes. In other regimes, however, it fails to stabilize Floquet eigenstates even when the resonance condition is satisfied; instead, the system can be driven into a maximally mixed state, as we demonstrate below. We attribute this failure to a Red transition process that competes with the Blue one in Fig.~\ref{process}(b). Owing to the cavity-spin interaction, the system can absorb energy from multiple drive photons and undergo a transition from $|u_-,0\rangle$ to $|u_+,1\rangle$, a process that is strongly suppressed at high driving frequencies but can be enhanced at intermediate frequencies. Subsequent photon loss then takes the system from $|u_+,1\rangle$ to $|u_+,0\rangle$, competing with the Blue process. The coexistence of the two processes thus causes the stabilization to fail.

We emphasize that the appearance of this competition depends on the Hamiltonian: it is not limited to our particular model, but neither is it guaranteed to occur in every Floquet model. A clear example is the exactly solvable rotating-field model of Ref.~\cite{ritterAutonomousStabilizationFloquet2025}: at any driving frequency, 
% the Fourier spectrum of its transition matrix elements contains only three frequencies, with the result that 
no driving frequency of the Red process satisfies the resonance condition at a positive cavity frequency, and the Blue process operates alone; we also explicitly show this case in Sec.~SM~3~\cite{supplement}. 

The system thus exhibits a rich variety of steady states, ranging from the maximally mixed state---corresponding to the complete breakdown of stabilization---to distinct target Floquet eigenstates, depending on parameters. To map out these steady states, we numerically simulate the master equation Eq.~\eqref{rhoME} over a range of driving frequencies $\omega$ and cavity frequencies $\Delta$. In the model studied, we find the steady state to be unique and independent of initial conditions.

To diagnose whether stabilization is successful, we use two complementary quantities. The first is the von Neumann entropy of the spin system, $S_v= -\text{Tr}[\rho_S\ln \rho_S]$, where $\rho_S=\text{Tr}_C[\rho]$ is the reduced density matrix after tracing out the cavity. If the spin is stabilized in a pure Floquet eigenstate, the entropy should be close to zero; conversely, a maximally mixed state yields $S_v=\ln 2$. The second diagnostic is the stroboscopic fidelity with respect to the Floquet eigenstates, defined as
\begin{equation}
    \langle P_\pm(nT) \rangle= \text{Tr}_S[ |u_\pm (nT)\rangle \langle u_\pm (nT)| \rho_S(nT) ].
\end{equation} A fidelity close to 1 indicates successful stabilization onto the target Floquet eigenstate.

\begin{figure}[!t]
    \centering
    \includegraphics[width=\linewidth]{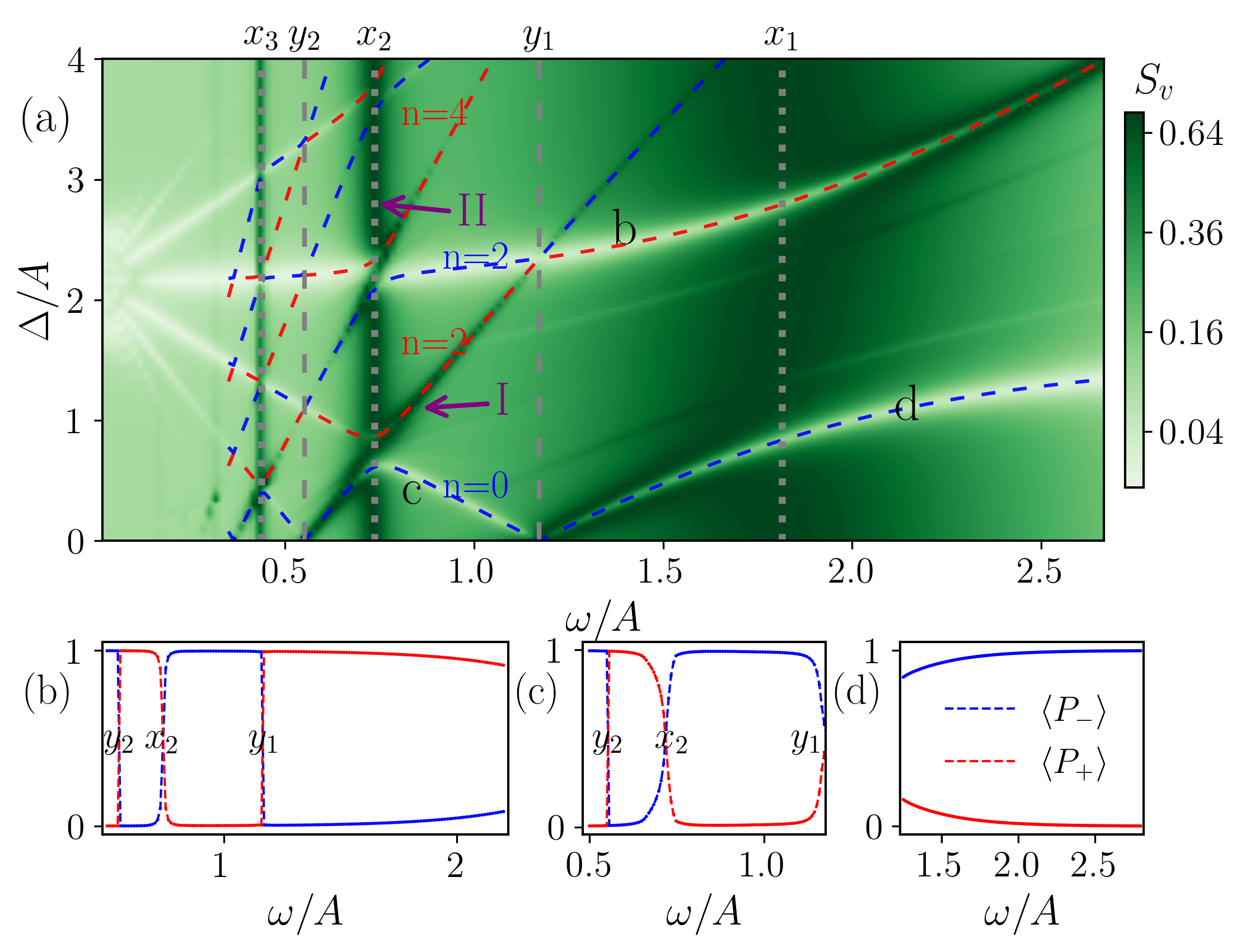}
    \caption{(a) Converged stroboscopic entropy of the two-level system as a function of cavity frequency $\Delta$ and driving frequency $\omega$. Blue dashed lines indicate $\Delta= \delta \varepsilon+n\omega$ and red dashed lines indicate $\Delta= -\delta \varepsilon+n\omega$. The dotted gray lines $\omega/A=x_i$ mark the frequency regions where the two matrix elements $\overline{M^2_{r/b}}$ cross (see Fig.~\ref{transM}), predicting high entropy. We classify the high-entropy regions into two categories: those distributed along the resonance conditions (Region I) and those vertical regions near the dotted lines (Region II), as indicated by the purple arrows. (b--d) Fidelities to the Floquet ground state (blue) or excited state (red) along the three low-entropy branches labeled (b--d) in panel (a). Values are converged after long-time evolution. Parameters: $A=60$, $\Omega=100$, $g=\kappa=0.05A$, and $\gamma=0.05\kappa$. The points $x_i$ (matrix-element crossovers) and $y_i$ (quasienergy degeneracies) are defined in Fig.~\ref{transM}.}
    \label{contourf}
\end{figure}
Figure~\ref{contourf}(a) shows the entropy landscape as a function of the cavity frequency $\Delta$ and the driving frequency $\omega$. The white branches represent parameter regions where the entropy approaches zero, while the dark green regions exhibit high entropy. To identify which Floquet eigenstate is stabilized, we select the three lowest-entropy branches [labeled (b--d) in Fig.~\ref{contourf}(a)] and plot the corresponding fidelity in panels (b--d), respectively. Along these branches, either the red or blue fidelity approaches unity, indicating that the spin stabilizes into different Floquet eigenstates depending on the driving frequency.

These high-fidelity white regions satisfy exactly one of the Red or Blue resonance conditions (except at quasienergy degeneracy, $\delta\varepsilon=0$). To demonstrate this, we overlay the resonance conditions for the two transition processes as blue and red dashed lines in Fig.~\ref{contourf}:
\begin{equation}
    \Delta= \pm \delta \varepsilon + n\omega,
    \label{Resonance}
\end{equation}
where $n$ is an even integer, the number of drive photons absorbed or emitted. We also remark that, owing to a dynamical parity symmetry of our system, which forbids odd-$n$ sidebands, only even $n$ contribute, see Sec.~SM~1.3 for details~\cite{supplement}. In the Red process, shown by red dashed lines, the cavity and spin absorb energy $n\omega$ from the drive to overcome the energy difference $\delta\varepsilon+\Delta$ between $|u_+,1\rangle$ and $|u_-,0\rangle$. In the Blue process, shown by blue dashed lines, the transition requires overcoming the smaller energy difference $\Delta-\delta\varepsilon$.

Within the dark regions, two distinct types of stabilization failure emerge:
\begin{itemize}
    \item (I) Some dark regions track the resonance-condition lines yet still produce high-entropy steady states (e.g., Region I in Fig.~\ref{contourf}(a)), demonstrating that satisfying the resonance condition is not sufficient for stabilization.
    \item (II) Other failure regions appear as narrow vertical windows in driving frequency that cut through the white branches (e.g., Region II), meaning that no choice of cavity frequency can stabilize the spin at those driving frequencies.
\end{itemize}

In the following, we use the Floquet Fermi golden rule (FGR) to show that this stabilization scheme is valid in the high-frequency regime~\cite{PhysRevB.104.134308}. When the high-frequency condition is not satisfied, however, the simple resonant-selection picture breaks down, and the two failure modes described above appear.

\subsection{Theoretical analysis in the high-frequency limit}

We first analyze the high-frequency limit, where the condition $\omega\gg A,\Omega$ holds and the FGR is applicable. Since $A \gg \kappa$ in our simulations, this also implies $2\pi/\omega \ll 1/\kappa$. In this regime, the system evolves for $\omega/(2\pi\kappa)$ driving periods on average before a photon loss event occurs. Accordingly, we can characterize the system dynamics via a non-Hermitian effective Hamiltonian $H_{\text{eff}}(t)=H_S(t)+(\Delta-i\kappa/2)a^\dagger a$, and we additionally treat the spin-cavity interaction $H_I$ as a perturbation. This perturbation induces transitions between product states $|u_\mu,n\rangle$ of Floquet eigenstates and photon number states (where $\mu=\pm$ and $n\in\mathbb{N}$). Taking the transition from the zero-photon state $|u_\nu,0\rangle$ to the single-photon state $|u_\mu,1\rangle$ as an example, the transition rate is derived as follows (details in the Sec.~SM~1.2~\cite{supplement}):
\begin{equation}
\begin{aligned}
\Gamma_{\mu,\nu} 
  &\approx \sum_n \left| M^{(n)}_{\mu,\nu} \right|^2  \frac{g^2\kappa}{(\varepsilon_{\mu}
-\varepsilon_{\nu}+\Delta+n\omega)^2+(\kappa/2)^2},   \\
\text{with}\, & M^{(n)}_{\mu,\nu}= \frac{1}{T}\int_0^T dt\langle  u_\mu(t) |\sigma^-| u_\nu(t)\rangle e^{-in\omega t}, 
\label{eq.FGR}
\end{aligned}
\end{equation}
where $M^{(n)}_{\mu,\nu}$ is the $n$-th Fourier component of the time-dependent transition matrix element between Floquet eigenstates $M_{\mu,\nu}(t)=\langle  u_\mu(t) |\sigma^-| u_\nu(t)\rangle$. Note that $M^{(n)}_{\mu,\nu}$ is non-zero only for even $n$ due to a dynamical parity symmetry~\cite{engelhardtDynamicalSymmetriesSymmetryProtected2021} (see Sec.~SM~1.3 for details). For the two-level system, we label the transition matrix elements $|\langle u_\pm(t)|\sigma^-|u_\mp(t)\rangle|$ as $M_{r/b}(t)$, corresponding to the Red and Blue processes in Fig.~\ref{process}, respectively.
%% 这段注释不要删除 The norm $| M^{(n)}_{\mu,\nu} |$ is independent of the choice of the initial-time $t_0$ (Floquet gauge). 

The transition rate in Eq.~\eqref{eq.FGR} is governed by the transition matrix elements and the energy denominator $\Delta E^{(n)}_{\mu,\nu}\equiv\varepsilon_{\mu}{-}\varepsilon_{\nu}{+}\Delta{+}n\omega$. The resonance condition of Eq.~\eqref{Resonance} is precisely the requirement that this denominator reduce to a small value $(\kappa/2)^2$ for some integer $n$. When such an $n$ exists, e.g., $\Delta E^{(n_0)}_{\mu,\nu}=0$ for a specific integer $n_0$,
the corresponding channel is resonant and dominates the sum over $n$ in Eq.~\eqref{eq.FGR}.
% when no such $n$ exists, the energy mismatch $\Delta E^{(n)}_{\mu,\nu}\equiv\varepsilon_{\mu}-\varepsilon_{\nu}+\Delta+n\omega$ remains of order $\omega$, so that in the high-frequency regime $\Delta E^{(n)}_{\mu,\nu}\gg\kappa$ and the channel is strongly suppressed.
% When the resonance condition of the Blue or Red process $\varepsilon_{\mu}-\varepsilon_{\nu}+\Delta+n_0\omega=0$ is satisfied for a specific integer $n_0$, 
It has a small denominator $(\kappa/2)^2$ in Eq.~\eqref{eq.FGR} while all other terms are off-resonant and can be neglected. The transition rate then reduces to
\begin{equation}
\Gamma_{\mu,\nu} \approx \frac{4g^2}{\kappa} \left| M^{(n_0)}_{\mu,\nu} \right|^2,
\label{eq:Pres}
\end{equation}
where the photon loss rate $\kappa$ sets the resonant linewidth.

By contrast, for a transition far from resonance, the minimum detuning $d_{\mu,\nu}\equiv\min_{n\in2\mathbb{Z}}|\Delta E^{(n)}_{\mu,\nu}|$ is much larger than the cavity linewidth, $d_{\mu,\nu}\gg\kappa$. Neglecting the linewidth in the denominators of Eq.~\eqref{eq.FGR}, we obtain
\begin{equation}
\begin{aligned}
\Gamma_{\mu,\nu}
\simeq g^2\kappa\sum_{n\in2\mathbb{Z}}\frac{|M^{(n)}_{\mu,\nu}|^2}{(\Delta E^{(n)}_{\mu,\nu})^2}
\leq\frac{g^2\kappa}{d_{\mu,\nu}^2}\sum_{n\in2\mathbb{Z}}|M^{(n)}_{\mu,\nu}|^2.
\end{aligned}
\label{eq:Pnonres}
\end{equation}
Compared with the resonant prefactor $4g^2/\kappa$ in Eq.~\eqref{eq:Pres}, the off-resonant prefactor is suppressed by a factor of order $g^2\kappa/d_{\mu,\nu}^2\sim g^2\kappa/\omega^2$. The relative transition rates therefore depend on both this spectral suppression and the corresponding Fourier weights.

% By contrast, in the non-resonant case, i.e., no integer $n$ satisfies the resonance condition, so all terms in Eq.~\eqref{eq.FGR} remain off-resonant. {In the high-frequency regime}, its denominators now scale as $\sim\omega\gg\kappa$, and the transition rates are of order
% \begin{equation}
% \Gamma_{\mu,\nu} \sim \frac{g^2\kappa}{\omega^2} \sum_n \left| M^{(n)}_{\mu,\nu} \right|^2,
% \label{eq:Pnonres}
% \end{equation}
% where we have used $g^2\kappa/(\Delta E^{(n)}_{\mu,\nu})^2\sim g^2\kappa/\omega^2$. 
% Multiple channels enable the system to absorb or emit multiple photons from the drive, facilitating Red or Blue processes\hz{I think Red or Blue can happen only when resonance condition is satisfied, no?}. 

Comparing Eqs.~\eqref{eq:Pres} and \eqref{eq:Pnonres}, a resonant transition is enhanced over a non-resonant one by a factor of order \(\sim(\omega/\kappa)^2\gg1\). {In principle, the sum over \(n\) in Eq.~\eqref{eq:Pnonres} could partially compensate for this resonant enhancement. In the high-frequency regime, however, appreciable Fourier components \(M_{\mu,\nu}^{(n)}\) are confined to small \(|n|\), while higher-order sidebands contribute negligibly. The sum therefore provides only a finite correction and cannot overcome the parametrically large resonant enhancement.} Consequently, the steady state in the high-frequency regime is overwhelmingly determined by the resonant transition channel. 

% For example\hz{which parameters are you considering? Without specifying the parameters, the following description seems to be repeating what we have said and does not provide more useful information}, when parameters are tuned to the Blue resonance away from quasienergy degeneracy, the Blue transition rate reaches $\Gamma_{-,+}\approx (4g^2/\kappa)|M^{(n_0)}_{-,+}|^2$, while the Red transition remains off-resonant with $\Gamma_{+,-}\ll \Gamma_{-,+}$. Provided the resonant Fourier component $M^{(n_0)}_{-,+}$ is nonzero, the system relaxes into the Floquet ground state via the Blue process. Tuning instead to the Red resonance stabilizes the Floquet excited state. 
% This resonant-selection mechanism underlies the high-fidelity white branches in Fig.~2(a). 

% At intermediate driving frequencies, however, multiple Fourier components can contribute appreciably to the transition rates, so resonance alone no longer suffices to determine the stabilized state. The resulting breakdown of the simple resonant-selection picture is analyzed in the next subsection.

We note in passing that in the extreme high-frequency limit, the Floquet Hamiltonian approaches the static form $H_F\to \frac{\Omega}{2}\sigma_z$, the Floquet states become eigenstates of $\sigma_z$, and the Red matrix element $M_r(t)$ vanishes ($\sigma^-$ annihilates the spin-down state), while the Blue matrix element $M_b(t)$ remains finite. As a result, only the Blue process survives, and only the branch corresponding to the Blue process retains high fidelity to Floquet ground state at large $\omega$, as shown in the region $\omega/A>2.5$ of Fig.~\ref{contourf}(a).

\subsection{Transition channels and the cooperation--competition mechanism}
At intermediate driving frequencies, $\omega\sim A,\Omega$, however, the system may fail to approach a pure Floquet eigenstate as evidenced by the dark regions I and II of Fig.~\ref{contourf}(a). 
% Away from the high-frequency limit, an exact theory is difficult to establish; the analysis below is therefore qualitative, separately estimating the two competing factors.
% This occurs since the summation over $n$ in the off-resonant contributions estimated by Eq.~\eqref{eq:Pnonres} can now compete with the resonant transition rate Eq.~\eqref{eq:Pres} in this frequency regime, even when the resonance condition $\varepsilon_{\mu}-\varepsilon_{\nu}+\Delta+n_0\omega=0$ is satisfied. 
% for a particular process, for example the Blue process. While this resonant channel is enhanced, the competing off-resonant channel, the Red process, is not fully negligible at moderate frequencies, so one must compare the resonant transition rate of Eq.~\eqref{eq:Pres} against the off-resonant contributions estimated by Eq.~\eqref{eq:Pnonres}. 
Two distinct factors are crucial to understand this failure: (i) the transition matrix elements $M^{(n)}_{\mu,\nu}$, which are determined solely by the Floquet states and their micromotion, and (ii) the resonance condition, which depends on the cavity frequency and the quasienergy spectrum. In the high-frequency limit, the resonance condition alone dictates the outcome of the stabilization protocol, but at intermediate frequencies the relative magnitudes of the transition matrix elements become increasingly important. In the following, we analyze both factors in detail.

\begin{figure}[!t]
    \centering
    \includegraphics[width=\linewidth]{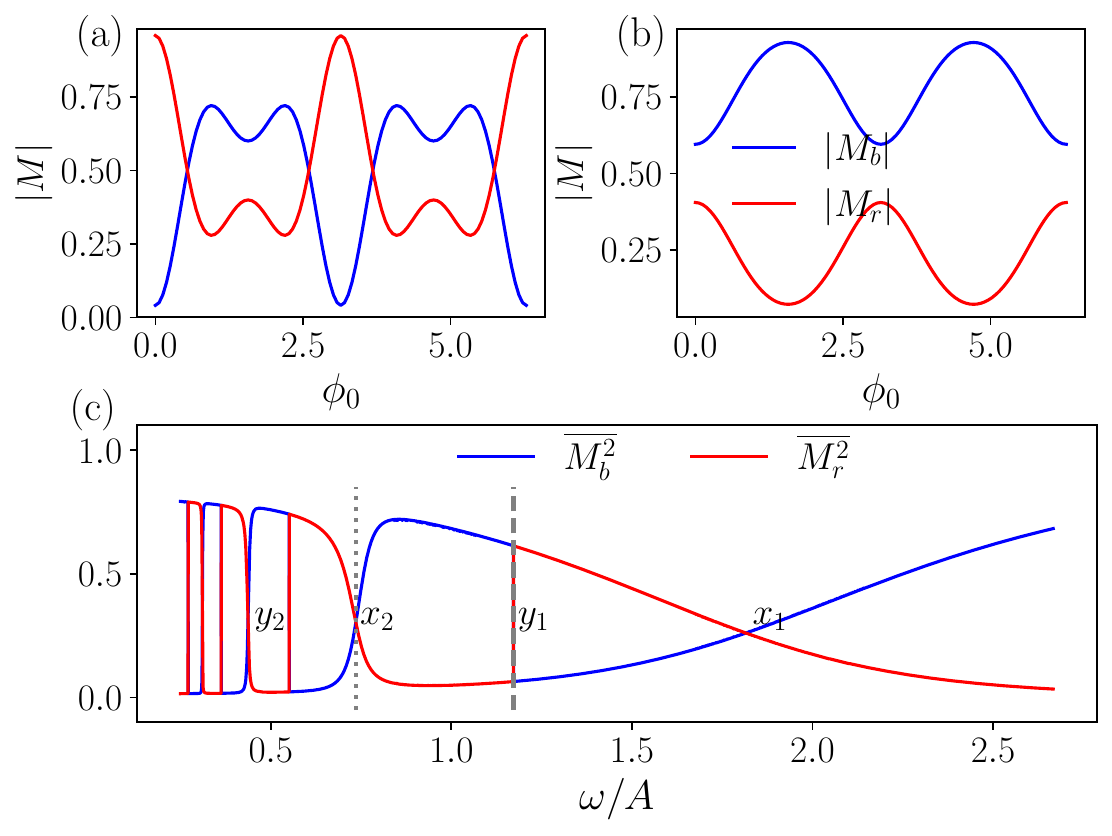}
    \caption{Transition matrix elements at different driving frequencies. (a,b) Instantaneous transition matrix elements $M_b=|\langle u_-(t)|\sigma^-|u_+(t)\rangle|$ (blue) and $M_r=|\langle u_+(t)|\sigma^-|u_-(t)\rangle|$ (red) as functions of the initial Floquet phase, at driving frequencies $\omega/A=0.736$ (a) and $\omega/A=1.17$ (b), marked by the dotted and dashed lines in (c), respectively. (c) Period-averaged squared matrix elements $\overline{M^2_{r/b}}$ versus driving frequency. By comparing with the fidelities in Fig.~\ref{contourf}(b-d), the dominant matrix element is found to indicate the stabilized Floquet state at intermediate driving frequencies, $\omega\sim A$ (e.g. $0.5<\omega/A<1.5$). The points $x_i$ label crossovers of the two matrix elements (predicting high-entropy regions), while $y_i$ label abrupt exchanges due to quasienergy degeneracy.}
    \label{transM}
\end{figure}

The instantaneous amplitudes $M_{r/b}(\phi)$, shown in Figs.~\ref{transM}(a,b) as functions of the Floquet phase $\phi=\omega t$, oscillate within each driving period. The relative strengths of the two processes depend on the driving frequency. At $\omega/A=0.736$ [Fig.~\ref{transM}(a)], the Red and Blue amplitudes alternate in dominance while remaining comparable, whereas at $\omega/A=1.17$ [Fig.~\ref{transM}(b)], one consistently dominates. Since the two processes favor opposite Floquet eigenstates, their comparable strengths near $\omega/A=0.736$ suggest that neither can dominate, preventing stabilization into a pure Floquet state.

% The relative magnitude of the two processes depends on the driving frequency. In Fig.~\ref{transM}(a) with $\omega/A=0.736$, the Red and Blue amplitudes repeatedly cross and alternate in dominance while remaining of comparable magnitude throughout the period, consistent with the near equality of their period-averaged weights. However, in Fig.~\ref{transM}(b) with $\omega/A=1.17$, they are well separated, with one consistently dominating the other. Since the two processes drive the system toward opposite Floquet eigenstates, their crossing and near equality in amplitude imply that neither can fully suppress the other, suggesting the failure of stabilization into a pure Floquet state near $\omega/A=0.736$.

To quantify this frequency dependence, we consider the period-averaged squared matrix elements,
\[
\overline{M_{r/b}^{2}}=\frac{1}{T}\int_0^T |M_{r/b}(t)|^2\,dt=\sum_n |M^{(n)}_{\pm,\mp}|^2.
\]
At intermediate driving frequencies, multiple Fourier components can contribute appreciably, making a direct comparison of individual transition channels cumbersome. By incorporating all Fourier components, the period-averaged squared matrix elements provide a simple phenomenological measure of the overall transition strengths in the two directions.

% To quantify this frequency dependence, we use the squared amplitudes averaged over one period, $\overline{M^2_{r/b}}=T^{-1}\int_0^T |M_{r/b}(t)|^2\,dt=\sum_n |M^{(n)}_{\pm,\mp}|^2$. At intermediate driving frequencies, many Fourier components are simultaneously nonzero, and comparing the individual components and opposite directions of $M^{(n)}_{\pm,\mp}$ would be complex and unclear. The period-averaged squared matrix element, which collects all Fourier components, instead provides a simple and transparent phenomenological measure of the total transition strength in each direction.

As shown in Fig.~\ref{transM}(c), $\overline{M^2_{r/b}}$ oscillates with $\omega$, and their relative magnitude exhibits two distinct types of crossings: continuous crossovers at $\omega/A=x_i$ (dotted gray lines) and abrupt exchanges at $\omega/A=y_i$ (dashed gray lines). For convenience, we label $x_i$ and $y_i$ from right to left, i.e., from high to low driving frequency. These points partition the frequency axis into alternating regimes in which either the Red or the Blue process dominates.

The two types of crossings have distinct physical origins. At $\omega/A=y_i$, the quasienergy difference vanishes, $\delta\varepsilon=0$, corresponding to an exact crossing of two quasienergy levels within the Floquet zone. Since quasienergies are defined modulo $\omega$, this crossing merely relabels the Floquet ground and excited states; the Red and Blue matrix elements swap their identities accordingly, producing the abrupt exchange $\overline{M^2_r}\leftrightarrow\overline{M^2_b}$. Crucially, the two matrix elements are still different at $y_i$: the one that was larger, as $\omega$ increases toward the crossing, remains larger afterward, only under a swapped label. The physical process that dominated before $y_i$ thus continues to dominate afterward, and the stabilization mechanism is not disrupted; we therefore expect near-zero entropy in the vicinity of $y_i$ for a resonance cavity frequency $\Delta$.

At $\omega/A=x_i$, the situation is qualitatively different: the quasienergy difference becomes comparable to the driving frequency, $\delta\varepsilon\approx\omega$, or to say the two quasienergies approach $+\omega/2$ and $-\omega/2$, respectively, within the first Floquet zone. It induces hybridization between Floquet states in adjacent zones and an avoided crossing of the quasienergies, producing a genuine crossover where $\overline{M^2_r}=\overline{M^2_b}$. Unlike at $y_i$, the two matrix elements here become equal, and no dominant channel is exhibited. Moreover, near $x_i$ the two processes can be simultaneously nearly resonant. Specifically, the cavity frequencies associated with the Blue process of driving photon $n$ and the Red process of driving photon $n+2$ are separated by $2|\omega-\delta\varepsilon|$, which becomes small when $\delta\varepsilon\approx\omega$. This is reflected by the approach of the red and blue dashed lines in Region II of Fig.~\ref{contourf}(a). At intermediate frequencies, the absence of a dominant channel together with this near-degeneracy of the two resonance conditions means that neither process can suppress the other and hence the stabilization fails for all cavity frequencies $\Delta$.

%% （不要删）图上的\phi_0 应改成\phi

The interplay of these two factors also explains the remaining failure mode, Region~I, whose dark branches track the resonance-condition lines in Fig.~\ref{contourf}(a). Along such a branch, the resonance condition is satisfied for one process while the dominant matrix element belongs to the other: within Region~I of Fig.~\ref{contourf}(a), the cavity is tuned to the Red resonance $\Delta=-\delta\varepsilon+2\omega$, yet this line crosses a frequency interval in which the Blue transition matrix element dominates ($\overline{M^2_b}>\overline{M^2_r}$). At moderate driving frequencies, the resonant enhancement factor $(\omega/\kappa)^2$ alone no longer determines the strongest transition channel; the disparity between the matrix elements can compensate for the lack of resonance. 
The off-resonant Blue process therefore competes with the resonant Red one, driving the system toward a mixed steady state with entropy approaching $\ln 2$. Thus, satisfying the resonance condition alone does not guarantee stabilization; the resonant channel must also dominate over competing transitions.

% In summary, at intermediate driving frequencies, cooperation between the resonance condition and the dominant matrix element yields the white branches, whose steady states are close to pure Floquet states, competition between the two yields the Region~I dark branches with mixed states, and Floquet-zone hybridization near $\omega/A=x_i$ yields Region~II, approaching the maximally mixed states.

\section{Multilevel generalization: spin-$S$ systems}

\begin{figure}
    \centering
    \includegraphics[width=0.95\linewidth]{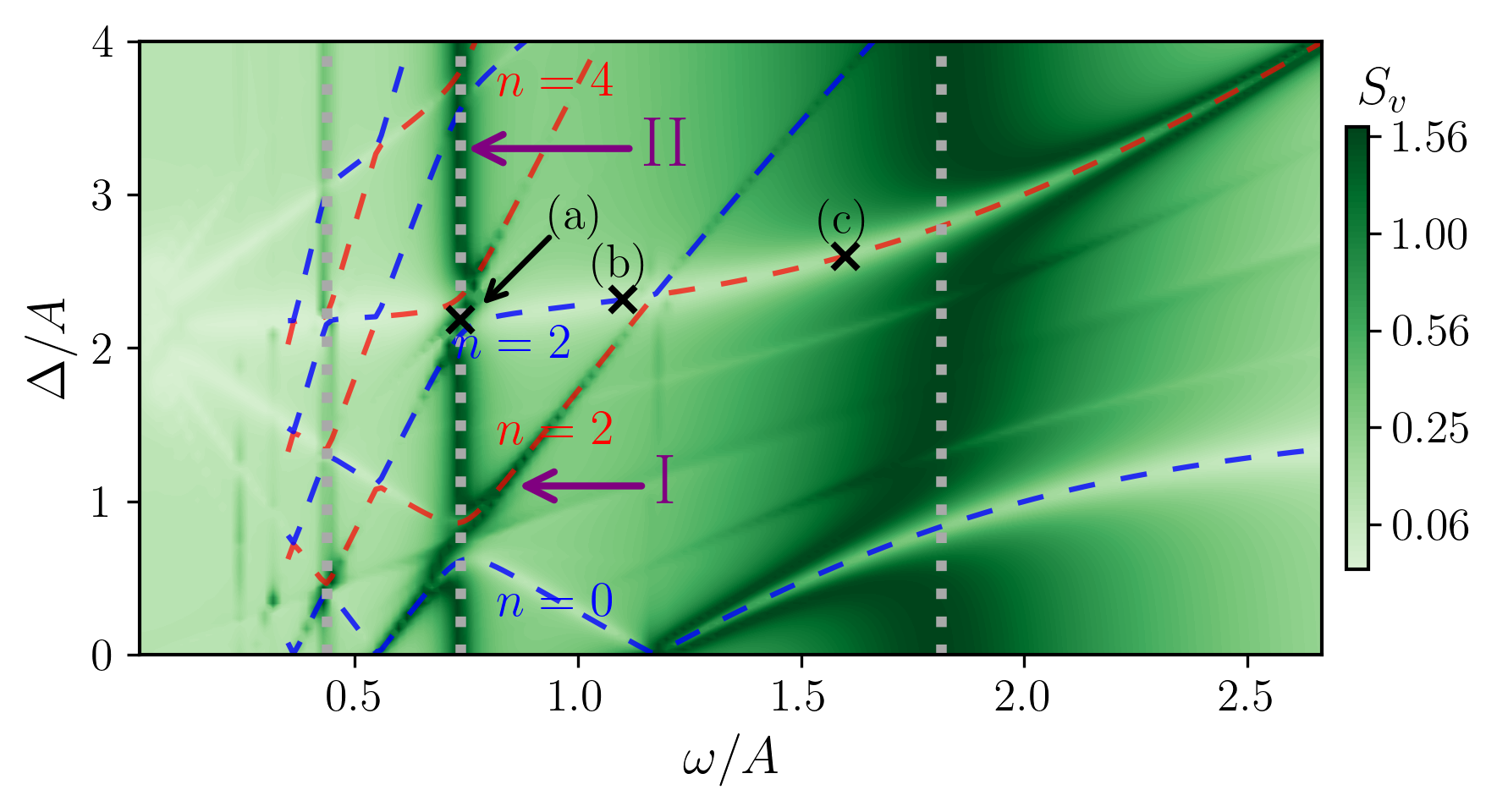}
    \caption{Entropy landscape of the five-level (\(S=2\)) system as a function of cavity frequency \(\Delta\) and driving frequency \(\omega\). The cooperation--competition framework of the two-level system persists in the multilevel case: low-entropy branches follow the sideband-resonance conditions, while two types of high-entropy regions appear. In Region~I, the leading transition pathway conflicts with the resonant pathway; in Region~II, Floquet-zone hybridization makes the competing transition channels comparable. Blue (red) dashed lines denote the Blue (Red) sideband-resonance conditions, \(\Delta=\pm\delta\varepsilon_F+n\omega\). Black arrows indicate the three parameter points (a--c) analyzed in Fig.~\ref{arrow}, and purple arrows mark Regions~I and~II. Parameters are the same as those in Fig.~\ref{contourf}.}
    \label{contourfd}
\end{figure}

We now extend our analysis to multilevel Floquet systems, where multiple transition pathways connect different Floquet states. Unlike the two-level case, stabilization requires these pathways to collectively direct population toward a unique target state. This raises a new question: under what conditions can such a globally preferred relaxation pathway emerge? We address this question using a driven spin-$S$ system with $d=2S+1$ levels, whose $\mathfrak {su}(2)$ structure allows us to characterize the transition pathways analytically.

As we show below, the $\mathfrak {su}(2)$ structure guarantees a uniformly spaced unfolded Floquet spectrum and constrains the cavity-induced transitions to a single chain with a uniform preferred direction. These properties allow us to establish the conditions for multilevel stabilization and explain why the same three regimes observed in the two-level system persist.

\subsection{Model and quasienergy structure}
We generalize the model of Eq.~\eqref{Fmodel} to a collective spin $S$ by replacing the Pauli operators with spin-$S$ operators, $\sigma_i/2\to S_i$:
\begin{equation}
 H_{S}(t)=-2A \cos(\omega t)S_{y}+\Omega S_{z} ,\,
 H_{I}=g(a^\dagger S^-  + aS^+ ) ,
\label{FmodelS}
\end{equation}
where $S_i$ ($i\in\{x,y,z\}$) are the spin-$S$ operators with $S_z|m\rangle=m|m\rangle$ ($m=-S,\dots,S$), $S^\pm$ are the spin ladder operators, and the cavity Hamiltonian $H_C=\Delta a^\dagger a$ is unchanged. As in the spin-1/2 case, we use the von Neumann entropy $S_v$ of the $d$-level system to characterize stabilization. The resulting entropy landscape in Fig.~\ref{contourfd} is qualitatively similar to that of the two-level system of Fig.~\ref{contourf}; in particular, the sideband-resonance condition and the high-entropy regions predicted for the two-level system agree well with those observed in the multilevel system. We explain this similarity below.

To begin with, we write the Hamiltonian as $H_S(t)=\vec{B}(t)\cdot\hat{\vec{S}}$, where $\vec{B}(t)$ is the time-dependent driving field. After a unitary rotation $P(t,t_0)$ into the micromotion frame, the Floquet Hamiltonian retains an analogous form $H_F(t_0)=\vec{B}_F(t_0)\cdot\hat{\vec{S}}$, depending only on the initial time. A further rotation $R(t_0)$ diagonalizes it to $H_F^\prime=B_z^\prime\hat{S}_z$, with eigenstates $|m\rangle$ ($m=-S,\dots,S$) related to the Floquet states by $|u_m(t)\rangle=P(t,t_0)R(t_0)|m\rangle$~\footnote{Since $\{S_z\}$ is a Cartan subalgebra of $\mathfrak{su}(2)$ by the Cartan conjugacy theorem, every element of $\mathfrak{su}(2)$ can be brought into $\{S_z\}$ by an adjoint transformation.}. We refer to the $\varepsilon_{F,m}$ as the unfolded Floquet spectrum, in contrast to the quasienergies $\varepsilon_i$ folded into the first Floquet zone. Crucially, the Floquet energies $\varepsilon_{F,m}=mB_z^\prime$ are uniformly spaced with $\delta\varepsilon_F=B_z^\prime$, identical to the spacing of the unfolded quasienergies in the spin-1/2 model. This equality is the key for the generalization to multilevel systems: it implies that all adjacent Floquet pairs in the unfolded spectrum satisfy the same form of resonance condition.

% At finite driving frequency, however, the quasienergies, folded into a single Floquet zone, no longer have uniform spacings. We label these folded quasienergies by $i=0,1,\dots,2S$ in ascending order, $\varepsilon_0<\varepsilon_1<\cdots<\varepsilon_{2S}$, within the chosen zone; the mapping between this sorted labeling and the ladder index $m$ is generally non-monotonic. Consequently, different adjacent pairs generally have different quasienergy spacings. Throughout this section, ``adjacent'' or ``neighboring'' states refer to pairs of states $|m\rangle\leftrightarrow|m-1\rangle$ of the unfolded spin ladder, unless explicitly stated otherwise.

After folding into the first Floquet zone, however, the quasienergies \(\varepsilon_i\), labeled in ascending order, generally have nonuniform spacings and a nonmonotonic correspondence with the ladder index \(m\). Throughout this section, ``adjacent'' or ``neighboring'' states refer to \(|m\rangle\) and \(|m-1\rangle\) in the unfolded ladder, unless explicitly stated otherwise.

\subsection{Resonance condition and sideband-resonance pathway}

Although the folded quasienergies are not equally spaced, the uniform spacing of the unfolded spectrum suffices to guarantee a simultaneous resonance condition. Consider two adjacent ladder states $|m\rangle$ and $|m-1\rangle$, whose folded quasienergies can be written as $\varepsilon_i=\varepsilon_{F,m}-\ell_m\omega$ and $\varepsilon_j=\varepsilon_{F,m-1}-\ell_{m-1}\omega$ with integer Floquet-zone shifts $\ell_m,\ell_{m-1}\in\mathbb{Z}$. Their difference therefore satisfies
\begin{equation}
 \delta \varepsilon_{i,j} - \delta \varepsilon_F = \ell\omega, \quad \ell = 0, \pm 1,
 \label{RC2}
\end{equation}
where $\ell\equiv\ell_{m-1}-\ell_m$ is restricted to $0,\pm1$ for adjacent ladder pairs, since folding shifts an adjacent pair of Floquet energies by at most one Floquet zone. 
Consequently, a single cavity frequency satisfying
$\Delta=\pm\delta\varepsilon_F+n\omega
$
can simultaneously satisfy the sideband-resonance conditions for all neighboring pairs, with appropriately shifted sideband indices.
% Consequently, a single sideband-resonance condition $\Delta=\pm\delta\varepsilon_{F}+n\omega$ simultaneously implies a resonance $\Delta=\pm\delta\varepsilon_{i,j}+n^\prime\omega$ on every link of the ladder, with a link-dependent integer $n^\prime=n\mp\ell$. This simultaneous resonance is the key property that makes multilevel stabilization possible.

We illustrate this property using the five-level system ($S=2$). At $\omega/A=1.41$, the folded quasienergy ordering coincides with that of the unfolded spin ladder, so all neighboring quasienergy levels are equally spaced by $\delta\varepsilon_F$\footnote{The folded quasienergy spectrum also possesses an antiunitary particle-hole symmetry \(\mathcal C=e^{-i\pi S_y}K\), where \(K\) denotes complex conjugation. This symmetry satisfies \(\mathcal C U(T)\mathcal C^{-1}=U(T)\) and enforces \(\varepsilon_i=-\varepsilon_{2S-i}\), implying \(\delta\varepsilon_{2S-i,2S-j}=\delta\varepsilon_{j,i}\). This symmetry is not required for the general simultaneous-resonance argument.}.
% At $\omega/A=1.41$, our numerical simulation suggests that $\delta\varepsilon_{4,3}$ and $\delta\varepsilon_{3,2}$ coincide with the two-level splitting $\delta\varepsilon_F$. The remaining two adjacent quasienergy differences are fixed by an antiunitary particle-hole symmetry $C=e^{-i\pi S_y}\mathcal{K}$ with $\mathcal{K}$ the complex conjugation, which leaves the one-period evolution operator invariant, $C U(T)C^{-1}=U(T)$, and thus pairs the quasienergies as $\varepsilon_{i}=-\varepsilon_{2S-i}$, and hence $\delta\varepsilon_{2S-i,2S-j}=\delta\varepsilon_{j,i}$. For $S=2$, this gives $\delta\varepsilon_{2,1}=\delta\varepsilon_{3,2}$ and $\delta\varepsilon_{1,0}=\delta\varepsilon_{4,3}$, so all four quasienergy differences become equal and the folded quasienergy ordering coincides with the unfolded one. This symmetry simplifies the $S=2$ example but is not required for the general simultaneous-resonance argument.
The Red and Blue sideband processes then form simple monotonic pathways. 
\begin{equation}
    \begin{matrix}
 \text{Red sideband: } & 0\to1\to2\to3\to4, \\
 &  \Delta=-\delta \varepsilon_F+n\omega, \\
 \text{Blue sideband: }& 4\to3\to2\to1\to0, \\
 &  \Delta=\delta \varepsilon_F+n\omega,
\end{matrix}
\end{equation}
where $n\in 2\mathbb{Z}$; as in the two-level system, the dynamical parity symmetry forbids odd-$n$ sidebands. 

% At other driving frequencies, the folding may reshuffle the level ordering; the resulting Red and Blue sideband pathways at representative driving frequencies are listed in Table~\ref{tab:pathway}. These examples illustrate that, although the folded ordering may vary, the sideband-resonance condition can always be satisfied simultaneously for all adjacent pairs. The red and blue dashed lines in Fig.~\ref{contourfd} represent precisely the resonance conditions of these Red and Blue sideband pathways.

At other driving frequencies, Floquet-zone folding can reshuffle the quasienergy ordering, leading to the nonmonotonic pathways listed in Table I. Nevertheless, these pathways remain opposite orientations of the same unfolded spin ladder, and a single cavity frequency can simultaneously satisfy the sideband-resonance conditions for all neighboring pairs.

\begin{table}[!t]
    \centering
    \caption{Red and Blue sideband pathways of the five-level ($S=2$) system at representative driving frequencies. Quasienergies are labeled $i=0,\ldots,4$ in ascending order within the chosen Floquet zone. The Blue pathway is the reverse of the Red one. All pathways follow the same unfolded spin ladder, $|2\rangle\rightleftharpoons|1\rangle\rightleftharpoons|0\rangle\rightleftharpoons|-1\rangle\rightleftharpoons|-2\rangle$, but their folded quasienergy labels vary with driving frequency. The frequencies $\omega/A=1.1$ and $1.6$ correspond to points (b) and (c) in Fig.~\ref{contourfd}, respectively, and are further analyzed in Fig.~\ref{arrow}.}
    \label{tab:pathway}
    \begin{tabular}{c|c|c}
        \hline\hline
        $\omega/A$ & Red sideband & Blue sideband \\
        \hline
        $1.0$& $4\to1\to2\to3\to0$ & $0\to3\to2\to1\to4$ \\
        $1.1$ & $0\to1\to2\to3\to4$ & $4\to3\to2\to1\to0$ \\
        $1.6$ & $3\to0\to2\to4\to1$ & $1\to4\to2\to0\to3$ \\
        \hline\hline
    \end{tabular}
\end{table}

\subsection{Leading transition pathway and uniqueness}

We now show that the cavity-spin interaction permits only a single bidirectional transition pathway, whose leading direction determines the unique stabilized Floquet state.

Under the unitary rotation $P(t,t_0)R(t_0)$, the interaction Hamiltonian $H_I$ can be formally written as
\[
H'_I=a^\dagger[g_1(t,t_0)S^- + g_2(t,t_0)S^+ + g_3(t,t_0)S^z ] + \text{h.c.}\,,
\]
where $g_i(t,t_0)$ ($i\in\{1,2,3\}$) are coefficients depending on $t$ and $t_0$. Closure of the $\mathfrak {su}(2)$ algebra under rotations gives this form. The $S^z$ term is diagonal in the $\{|m\rangle\}$ basis and therefore does not induce transitions. The remaining $S^\pm$ terms give rise to two opposite transition paths linking adjacent Floquet states, $|S\rangle\rightleftharpoons|S-1\rangle\rightleftharpoons\cdots\rightleftharpoons|-S\rangle$, which together form a single bidirectional transition pathway.

The leading transition pathway is defined by selecting, at each pair of adjacent Floquet levels, the direction with the larger time-averaged transition matrix element. When the resonance condition is fulfilled, this leading transition pathway, which possesses a single terminal node, determines the stabilized Floquet eigenstate. By contrast, if the transition graph possesses multiple terminal nodes, the system might not relax to a single Floquet eigenstate.

We now prove that the direction of the leading transition pathway is unique. The time-averaged squared transition matrix element from $|S,m\rangle$ to $|S,m-1\rangle$ (driven by $S^-$) reads
\[
\overline{M^2_{(m-1,m)}}
=\frac{1}{T}\int^{t_0+T}_{t_0} dt\, |g_1(t,t_0)|^2(S+m)(S-m+1),
\]
where the Clebsch--Gordan (CG) factor $(S+m)(S-m+1)=|\langle S,m-1|S^-|S,m\rangle|^2$ originates from the spin algebra. The reverse transition from $|S,m-1\rangle$ to $|S,m\rangle$ (driven by $S^+$) has matrix element $\langle S,m|S^+|S,m-1\rangle=\langle S,m-1|S^-|S,m\rangle^*$ by hermiticity, so its squared modulus carries the \emph{same} CG factor $(S+m)(S-m+1)$. Consequently,
\[
\frac{\overline{M^2_{(m-1,m)}}}{\overline{M^2_{(m,m-1)}}}
=
\frac{\int_{t_0}^{t_0+T} dt\, |g_1(t,t_0)|^2}{\int_{t_0}^{t_0+T} dt\, |g_2(t,t_0)|^2},
\]
where the CG factor cancels. The ratio is therefore independent of $m$: at every adjacent pair, the same direction is preferred, so the leading transition pathway is a globally oriented chain with a single terminal node. When the sideband-resonance pathway coincides with this leading transition pathway, population is funneled along the chain and the system is stabilized into the unique Floquet state at its terminal node.

\begin{figure}[!t]
    \centering
    \includegraphics[width=0.95\linewidth]{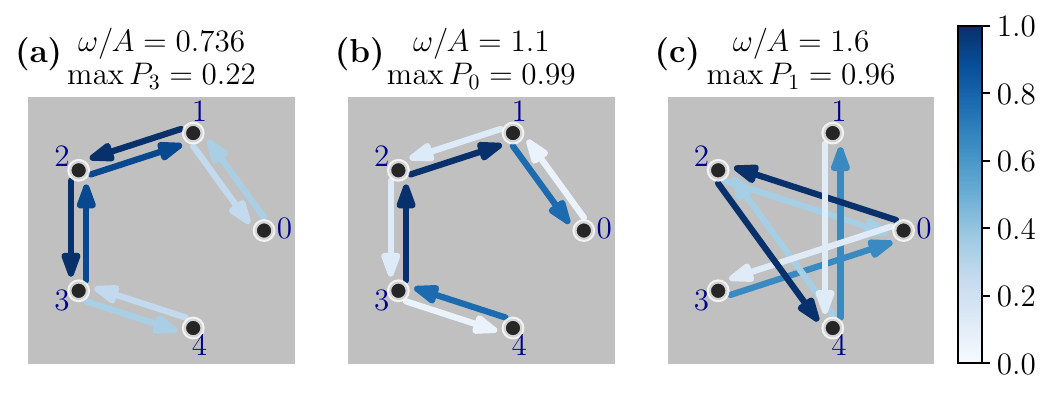}
    \caption{(a--c) Time-averaged squared transition matrix elements $\overline{M^2}$ of the five-level ($S=2$) system, visualized as directed graphs, at the three parameter points marked by arrows in the entropy landscape of Fig.~\ref{contourfd} ($\omega/A=0.736,1.1,1.6$). Each node represents a quasienergy level, numbered in ascending order within the chosen Floquet zone; a pair of oppositely directed arrows connects every adjacent pair, with the color depth encoding $\overline{M^2}$. In panels (b) and (c), one direction dominates at every adjacent pair of the unfolded quasienergies, so the transition graph possesses a single terminal node, which becomes the stabilized Floquet eigenstate when the sideband-resonance condition is satisfied. In panel (a), near the Floquet-zone hybridization region, the two directions become equal and this uniqueness is destroyed. The maximal fidelity $P_i$ to the Floquet eigenstate of the $i$-th quasienergy is annotated in each panel. Parameters $A$ and $\Omega$ are the same as those in Fig.~\ref{contourf}.}
    \label{arrow}
\end{figure}

Fig.~\ref{arrow} presents the time-averaged squared transition matrix elements $\overline{M^2}$ of the five-level system, visualized as directed graphs, at the three parameter points (a--c) {marked by black crosses in the entropy landscape of Fig.~\ref{contourfd}.} Each node represents a quasienergy level, and a pair of oppositely directed arrows connects every adjacent pair, {with the color depth encoding the magnitude of $\overline{M^2}$: darker arrows correspond to stronger transitions.} In panels (b) and (c), one arrow in each pair is markedly darker than the other, identifying a well-defined leading direction; the leading transition pathway is thus a globally oriented chain with a single terminal node, in agreement with the proof above. In panel (a), by contrast, located near the Floquet-zone hybridization region ($\delta\varepsilon_F\approx\omega$), hybridization drives a character swap between Floquet states that equalizes the oppositely directed transitions, so that no leading transition pathway exists and the uniqueness breaks down.

\subsection{Numerical verification: three regimes}

The sideband-resonance condition established in Sec.~III.B and the unique leading transition pathway established in Sec.~III.C together open the possibility of stabilizing the system into a single Floquet eigenstate; whether this stabilization succeeds and which eigenstate is selected depend on the interplay between these two ingredients. We now verify numerically that the same three regimes---cooperative, competitive, and near-degenerate---emerge in the multilevel system. To identify the stabilized Floquet eigenstate quantitatively, we compute the fidelity of each quasienergy eigenstate, defined as
\begin{equation}
    \langle P_i(nT) \rangle = \text{Tr}_S\left[  |u_i(nT) \rangle \langle  u_i(nT) | \rho_S(nT) \right],
\end{equation}
where $|u_i\rangle$ ($i\in\{0,1,\dots,4\}$) are the eigenstates corresponding to the quasienergies $\varepsilon_i$ of the five-level system. 

The maximal fidelity among all Floquet states at each of the three points of Fig.~\ref{arrow} is annotated in the corresponding panel, with the cavity frequency tuned to the Blue sideband resonance at points (a) and (b) and to the Red sideband resonance at point (c).
% In the entropy landscape of Fig.~\ref{contourfd}, points (b) and (c) lie on the white low-entropy branches, consistent with their high fidelities, while point (a) lies in a dark high-entropy region, with $P_2=0.27$ approaching the maximally mixed value $1/5$.

In the frequency region $\omega/A<2$, the leading transition pathway inferred from Fig.~\ref{arrow} dictates the steady Floquet state whenever it aligns with the sideband-resonance condition. The transition graphs of panels (b) and (c) possess terminal nodes $0$ and $1$, respectively, matching exactly the Floquet eigenstates selected by the maximal fidelities. The change of the terminal node between the two frequencies reflects a reversal of the leading direction as the driving frequency varies, analogous to the direction switching at $y_i$ in the two-level case. These are cooperative cases: the leading transition pathway aligns with one of the sideband-resonance pathways listed in Table~\ref{tab:pathway}, leading to nearly vanishing entropy.

When the sideband-resonance pathway and the leading transition pathway conflict, the Red and Blue sideband processes become comparable, producing the Region~I dark bands in Fig.~\ref{contourfd}: these regions exhibit finite entropy despite satisfying the sideband-resonance condition. The Region~II dark bands, located at $\delta\varepsilon_F\approx\omega$, instead arise from the Floquet-zone hybridization analyzed in Fig.~\ref{arrow}(a), in full analogy with the near-degenerate regime of the two-level case.

In summary, although the transition processes in the multilevel system are more intricate than in the two-level case, the underlying cooperation--competition mechanism and the resulting entropy structure remain essentially the same.

\section{Discussion}

In this work, we have shown that cavity-assisted stabilization of Floquet eigenstates can fail even when the cavity satisfies a sideband-resonance condition. This breakdown originates from competition between cavity-mediated transition-dissipation pathways that favor opposite Floquet eigenstates. In the high-frequency regime, resonant transitions are enhanced by a factor of order $(\omega/\kappa)^2$, typically suppressing competing off-resonant channels and recovering the stabilization mechanism of Ref.~\cite{ritterAutonomousStabilizationFloquet2025}. At intermediate driving frequencies, however, the transition matrix elements become equally important: sufficiently strong off-resonant processes can compete with resonant ones, preventing stabilization into a pure Floquet state. 

We further demonstrate that this framework extends to multilevel spin-$S$ systems. Their uniformly spaced unfolded Floquet spectra allow simultaneous sideband resonance between neighboring levels, while the $\mathfrak{su}(2)$ algebra establishes a uniform preferred direction for the period-averaged transitions. Together, these properties give rise to the same three regimes observed in the two-level system.

% The Supplementary Material extends these results in two directions. First, we show that the mechanism persists in generic quasiperiodically driven systems: when the dominant transition matrix element is compatible with the resonance condition, the system can be stabilized to an eigenstate of the quasiperiodically driven system with near-zero entropy, whereas a mismatch leads to a maximally mixed state. Second, we demonstrate that the stabilization strategy is robust against weak dephasing noise, which is ubiquitous in experimental setups.

More broadly, separating the effects of resonance selection and transition-matrix-element strengths provides a practical diagnostic for identifying parameter regimes where cavity-assisted stabilization succeeds or fails. Extending this framework to driven many-body systems coupled to dissipative cavities is a natural next step~\cite{steinfadtDissipationAssistedPreparationFloquetLaughlin2026, ramanDissipativeStabilizationFloquetEngineered2026}. In such systems, the proliferation of competing transition pathways and increasingly dense quasienergy spectra may give rise to richer stabilization phenomena, potentially including nonequilibrium steady-state phase transitions.
\section*{Acknowledgments.}
This work is supported by the National Natural Science Foundation of China (Grants No. 12234002, 12474214, 12474486, and 92250303), by the National Key Research and Development Program of China (Grant No. 2024YFA1612101), and by Quantum Science and Technology-National Science and Technology Major Project
(No. 2024ZD0301800), and by “The Fundamental Research Funds for the Central Universities, Peking University” and “High-performance Computing Platform of Peking University”.

\bibliography{references}
\newpage

\let\addcontentsline\oldaddcontentsline
	\cleardoublepage
	\onecolumngrid

\begin{center}
\textbf{\large{\textit{Supplementary Material} \\ \smallskip
	Resonance and Transition-Channel Competition in Cavity-Assisted Floquet-State Stabilization }}\\
\end{center}

\onecolumngrid
	\renewcommand{\thefigure}{S\arabic{figure}}
        \setcounter{figure}{0}
    \renewcommand{\thesection}{SM\;\arabic{section}}
	\setcounter{section}{0}
	\renewcommand{\theequation}{S.\arabic{equation}}
        \setcounter{equation}{0}
    \renewcommand{\thesubsection}{\arabic{section}.\arabic{subsection}}
	\setcounter{section}{0}
    \tableofcontents

\section{Floquet Fermi golden rule and symmetry-protected selection rules}

In this section, we derive the Floquet Fermi golden rule (FGR) used in the main text directly from the microscopic model of Eq.~\eqref{Fmodel}, and then show how a dynamical parity symmetry restricts the allowed Floquet sideband channels. We set \(\hbar=1\) throughout.
Before presenting the derivation, we summarize the underlying physical process. Population transfer between two Floquet eigenstates proceeds in two steps. First, the spin--cavity interaction coherently creates a cavity photon while coupling \(|u_\nu,0\rangle\) to \(|u_\mu,1\rangle\). This excitation is reversible, since the photon can be reabsorbed by the spin. Irreversible transfer occurs only when the cavity photon is lost, producing the quantum jump
$|u_\mu,1\rangle\rightarrow |u_\mu,0\rangle .
$
Photon loss therefore converts the coherent spin--cavity transition into an irreversible transition between Floquet states.

\subsection{The closed-system Floquet Fermi golden rule}

We first neglect cavity dissipation and determine the transitions induced by the spin--cavity coupling. Treating \(H_I\) perturbatively, we derive the transition amplitudes between Floquet states and show that they are controlled by the quasienergy differences and the Fourier components of the cavity-coupling matrix elements. This provides the closed-system Floquet Fermi golden rule, before we incorporate photon loss in the next subsection.

We work in the frame in which the model takes the periodic form of Eq.~\eqref{Fmodel}. Following the rotating-frame method of Ref.~\cite{pub.1145280638}, we introduce a periodic operator $K(t+T)=K(t)$ in terms of which the Floquet Hamiltonian is
\begin{equation}
    H_F = e^{iK(t)}\!\left[H_S(t)-i\frac{d}{dt}\right]e^{-iK(t)}.
    \label{eq:SM-HF}
\end{equation}
In the rotating frame $|\psi'(t)\rangle=e^{iK(t)}|\psi(t)\rangle$, the unperturbed spin therefore obeys $i\partial_t|\psi'\rangle=H_F|\psi'\rangle$ with a time-independent Hamiltonian. We write the Floquet eigenstates in this frame as
\begin{equation}
    |u'_\mu(t)\rangle=e^{iK(t)}|u_\mu(t)\rangle,
    \label{eq:SM-rotstate}
\end{equation}
where $|u_\mu(t+T)\rangle=|u_\mu(t)\rangle$ are the lab-frame Floquet modes and $H_F$ has quasienergies $\varepsilon_\mu$.

To set up perturbation theory we need the unperturbed propagator, which factorizes in terms of the micromotion operator:
\begin{equation}
    U_S(t,t_0)=e^{-iK(t)}e^{-iH_F(t-t_0)}e^{iK(t_0)},
    \label{eq:SM-US}
\end{equation}
which acts on a Floquet mode according to
\begin{equation}
    U_S(t,t_0)|u_\nu(t_0)\rangle=e^{-i\varepsilon_\nu(t-t_0)}|u_\nu(t)\rangle.
\end{equation}
Including the cavity, whose free evolution commutes with the spin, the full unperturbed propagator is
\begin{equation}
    \mathcal U_0(t,t_0)=U_S(t,t_0)\,e^{-i\Delta a^\dagger a\,(t-t_0)}.
    \label{eq:SM-U0}
\end{equation}

The perturbation that drives the transition is the spin--cavity coupling $V(t)=e^{iK(t)}H_Ie^{-iK(t)}$ in the rotating frame. For the transition $|\nu,0\rangle\to|\mu,1\rangle$, which raises the photon number by one, only the $a^\dagger\sigma^-$ term contributes. Its matrix element between rotating-frame Floquet states on the same section is
\begin{equation}
    \langle u'_\mu(t),1|V(t)|u'_\nu(t),0\rangle
    = g\langle u_\mu(t)|\sigma^-|u_\nu(t)\rangle
    \equiv g\,M_{\mu,\nu}(t),
    \label{eq:SM-Mt}
\end{equation}
where Eq.~\eqref{eq:SM-rotstate} cancels the micromotion factors. The matrix element is periodic, $M_{\mu,\nu}(t+T)=M_{\mu,\nu}(t)$, and admits the Fourier expansion
\begin{equation}
    M_{\mu,\nu}(t)=\sum_n M^{(n)}_{\mu,\nu}e^{in\omega t},
    \qquad
    M^{(n)}_{\mu,\nu}=\frac{1}{T}\int_0^T dt\,M_{\mu,\nu}(t)\,e^{-in\omega t},
    \label{eq:SM-Fourier}
\end{equation}
whose $n$-th harmonic describes the exchange of $n$ drive photons.

We now compute the transition amplitude to first order in $g$ using the full unperturbed propagator~\eqref{eq:SM-U0}. Starting from $|u_\nu(t_0),0\rangle$, the first-order Dyson amplitude for reaching $|u_\mu(t_0+\tau),1\rangle$ is
\begin{equation}
    c_{\mu,\nu}(\tau,t_0)
    =-i\int_{t_0}^{t_0+\tau}dt'\,
    \langle u_\mu(t_0+\tau),1|\,\mathcal U_0(t_0+\tau,t')H_I
    \mathcal U_0(t',t_0)|u_\nu(t_0),0\rangle .
    \label{eq:SM-cdyson}
\end{equation}
The two free propagations act on the matrix element as
\begin{equation}
    \mathcal U_0(t',t_0)|u_\nu(t_0),0\rangle
    =e^{-i\varepsilon_\nu(t'-t_0)}|u_\nu(t'),0\rangle,
\end{equation}
because the zero-photon state carries no cavity phase, and
\begin{equation}
    \langle u_\mu(t_0+\tau),1|\mathcal U_0(t_0+\tau,t')
    =e^{-i(\varepsilon_\mu+\Delta)(t_0+\tau-t')}
    \langle u_\mu(t'),1|,
\end{equation}
using Eq.~\eqref{eq:SM-US} for the spin and the free cavity evolution for the photon. With $s=t'-t_0$ and Eq.~\eqref{eq:SM-Mt}, we obtain
\begin{equation}
    c_{\mu,\nu}(\tau,t_0)
    =-ig\int_0^\tau ds\,
    e^{-i(\varepsilon_\mu+\Delta)(\tau-s)}
    e^{-i\varepsilon_\nu s}\,M_{\mu,\nu}(s+t_0).
    \label{eq:SM-c1}
\end{equation}

We now evaluate this integral explicitly and expose its multiphoton structure. Inserting the Fourier expansion in the form $M_{\mu,\nu}(s+t_0)=\sum_n M^{(n)}_{\mu,\nu}e^{in\omega t_0}e^{in\omega s}$, and defining the detuning associated with the $n$-th channel
\begin{equation}
    \Delta E^{(n)}_{\mu,\nu}\equiv\varepsilon_\mu-\varepsilon_\nu+\Delta+n\omega .
\end{equation}
The integral of each Fourier component is then
\begin{equation}
    c_{\mu,\nu}(\tau,t_0)
    =-ig\,e^{-i(\varepsilon_\mu+\Delta)\tau}
    \sum_n M^{(n)}_{\mu,\nu}e^{in\omega t_0}
    \frac{e^{i\Delta E^{(n)}_{\mu,\nu}\tau}-1}{i\Delta E^{(n)}_{\mu,\nu}}.
    \label{eq:SM-cn}
\end{equation}
Equation~\eqref{eq:SM-cn} describes reversible excitation of a discrete one-photon level. To resolve the individual Fourier channels, we average the excitation probability over the stroboscopic section $t_0$; the factors $e^{i(n-m)\omega t_0}$ then impose $\delta_{nm}$ and remove all interference between different channels:
\begin{equation}
    \frac{1}{T}\int_0^Tdt_0\,|c_{\mu,\nu}(\tau,t_0)|^2
    =g^2\sum_n|M^{(n)}_{\mu,\nu}|^2\,
    \frac{4\sin^2(\Delta E^{(n)}_{\mu,\nu}\tau/2)}
         {(\Delta E^{(n)}_{\mu,\nu})^2}.
\end{equation}
The long-time limit formally identifies the Floquet resonance conditions. We define the following expression in the distributional sense,
\begin{equation}
    \Gamma^{\mathrm{cl}}_{\mu,\nu}
    \equiv\lim_{\tau\to\infty}\frac{1}{\tau}\,
    \frac{1}{T}\int_0^Tdt_0\,|c_{\mu,\nu}(\tau,t_0)|^2,
\end{equation}
and use the identity
\begin{equation}
    \lim_{\tau\to\infty}\frac{1}{\tau}\frac{4\sin^2(x\tau/2)}{x^2}=2\pi\,\delta(x),
\end{equation}
which gives
\begin{equation}
    \Gamma^{\mathrm{cl}}_{\mu,\nu}
    =2\pi g^2\sum_n|M^{(n)}_{\mu,\nu}|^2\,
    \delta\!\left(\varepsilon_\mu-\varepsilon_\nu+\Delta+n\omega\right).
    \label{eq:SM-FGRclosed}
\end{equation}
This is the formal closed-system Floquet Fermi golden rule: energy is conserved modulo $n$ photons of the drive, and the $n$-th channel contributes only on the resonance surface $\Delta E^{(n)}_{\mu,\nu}=0$, with a strength set by the Fourier component $M^{(n)}_{\mu,\nu}$~\cite{PhysRevB.104.134308}. For a single lossless cavity mode there is no continuum, so away from an exact resonance the one-photon amplitude merely oscillates and no irreversible population transfer occurs; the $\delta$-function in Eq.~\eqref{eq:SM-FGRclosed} marks the envelope of allowed channels rather than a finite, measurable rate. A finite rate requires a finite cavity linewidth, to which we now turn.

\subsection{Floquet Fermi golden rule in the open quantum system with weak dissipation}

We now switch on the weak photon loss of the real cavity. Our goal is to show that dissipation turns the formal $\delta$-function rule of Eq.~\eqref{eq:SM-FGRclosed} into the finite Lorentzian rate of Eq.~\eqref{eq.FGR} used in the main text, and that it does so through an explicit physical process---the emission of a photon out of the cavity.

To count completed transitions with photon loss, we unravel the open-cavity dynamics into quantum trajectories~\cite{Dalibard1992,Plenio1998}. The total system obeys the Lindblad master equation
\begin{equation}
    \partial_t\rho=-i[H_S(t)+H_C+H_I,\rho]+\kappa\,\mathcal D_a[\rho]+\gamma\,\mathcal D_{\sigma^-}[\rho],
\end{equation}
with $\mathcal D_a[\rho]=a\rho a^\dagger-\tfrac12\{a^\dagger a,\rho\}$. We neglect intrinsic spin relaxation ($\gamma=0$) in the following derivation and treat $H_I$ to first order. On an individual trajectory, the system either loses a photon---a quantum jump---or evolves while no photon is detected. The evolution between two jumps, called the conditional or no-jump evolution, is generated by $H_{\mathrm{eff}}(t)+H_I$, where the unperturbed non-Hermitian Hamiltonian is
\begin{equation}
    H_{\mathrm{eff}}(t)=H_S(t)+\left(\Delta-i\frac{\kappa}{2}\right)a^\dagger a,
    \label{eq:SM-Heff}
\end{equation}
whereas a loss event acts through the jump operator $J=\sqrt{\kappa}\,a$, which maps $|\mu,1\rangle$ to $\sqrt{\kappa}\,|\mu,0\rangle$. The norm of the unnormalized conditional state equals the probability that no jump has occurred; therefore its rate of decrease is exactly the rate of jumps, i.e. of completed transitions. We compute this rate in three steps: the conditional amplitude between jumps, its long-time value, and the norm loss due to jumps.

In the first step, only the single-photon component of the unperturbed propagator~\eqref{eq:SM-U0} is modified, its energy acquiring the imaginary part $-i\kappa/2$:
\begin{equation}
    \begin{aligned}
    \langle u_\mu(t),1|\widetilde{\mathcal U}_0(t,t')
    &=e^{-i(\varepsilon_\mu+\Delta)(t-t')}e^{-\kappa(t-t')/2}
      \langle u_\mu(t'),1|,\\
    \widetilde{\mathcal U}_0(t',t_0)|u_\nu(t_0),0\rangle
    &=e^{-i\varepsilon_\nu(t'-t_0)}|u_\nu(t'),0\rangle ,
    \end{aligned}
    \label{eq:SM-eig}
\end{equation}
where $\widetilde{\mathcal U}_0$ is the damped propagator generated by $H_{\mathrm{eff}}$ and the second line is unchanged because $a^\dagger a|u_\nu,0\rangle=0$: a state with no photon cannot lose one. $H_{\mathrm{eff}}$ is a normal operator (its Hermitian parts act on different subsystems and commute), so the same complex eigenvalues appear in bra and ket matrix elements.

We now repeat the first-order calculation of Eq.~\eqref{eq:SM-c1} with the damped propagator, replacing $\mathcal U_0$ by $\widetilde{\mathcal U}_0$. This gives the conditional amplitude for finding one photon in the cavity at time $t_0+\tau$,
\begin{equation}
    c_{\mu,\nu}(\tau,t_0)
    =-ig\int_0^\tau ds\,
    e^{-i(\varepsilon_\mu+\Delta)(\tau-s)}
    e^{-\kappa(\tau-s)/2}
    e^{-i\varepsilon_\nu s}\,M_{\mu,\nu}(s+t_0).
    \label{eq:SM-cdamp}
\end{equation}
The placement of the damping factor is the crucial point: $e^{-\kappa(\tau-s)/2}$ acts only over the interval $[t',t_0+\tau]$ \emph{after} the photon has been created in the cavity at time $t'=t_0+s$. A photon cannot leak out before it exists. Damping the amplitude over the whole interval instead would amount to unphysical photon loss prior to the excitation and is precisely the error that prevents a naive treatment from reaching the rate form derived below.

In the second step, we evaluate this integral explicitly and take the long-time limit $\tau\gg\kappa^{-1}$, which will expose the Lorentzian line shape of each channel. Inserting the Fourier expansion of $M_{\mu,\nu}$ and integrating each harmonic gives
\begin{equation}
    \begin{aligned}
    c_{\mu,\nu}(\tau,t_0)
    =&-ig\,e^{-i(\varepsilon_\mu+\Delta)\tau}e^{-\kappa\tau/2}\times\sum_n M^{(n)}_{\mu,\nu}e^{in\omega t_0}
    \frac{e^{[i\Delta E^{(n)}_{\mu,\nu}+\kappa/2]\tau}-1}
         {i\Delta E^{(n)}_{\mu,\nu}+\kappa/2}.
    \end{aligned}
    \label{eq:SM-c2}
\end{equation}
Multiplying out the numerator gives two terms: one proportional to $e^{-i(\varepsilon_\nu-n\omega)\tau}$ and one proportional to $e^{-i(\varepsilon_\mu+\Delta)\tau}e^{-\kappa\tau/2}$. The second one vanishes for $\tau\gg\kappa^{-1}$, leaving the quasi-stationary conditional amplitude
\begin{equation}
    c^{\mathrm{ss}}_{\mu,\nu}(\tau,t_0)\simeq
    -\sum_n\frac{g\,M^{(n)}_{\mu,\nu}e^{in\omega t_0}
                 e^{-i(\varepsilon_\nu-n\omega)\tau}}
                 {\Delta E^{(n)}_{\mu,\nu}-i\kappa/2}.
    \label{eq:SM-c3}
\end{equation}
Each Fourier harmonic thus contributes with a Lorentzian denominator centered at its resonance $\Delta E^{(n)}_{\mu,\nu}=0$ and carries the stroboscopic phase $e^{in\omega t_0}$ of the initial section.

In the third step, we calculate the rate at which the photon in $|u_\mu,1\rangle$ escapes, leaving the spin in $|u_\mu\rangle$. The probability of occupying this one-photon state before a jump is $|c_{\mu,\nu}(\tau,t_0)|^2$, to leading order in $g$. The probability density of a photon-loss event follows from the norm loss,
\begin{equation}
    -\frac{d}{dt}\langle\psi(t)|\psi(t)\rangle
    =\langle\psi(t)|\,iH_{\mathrm{eff}}-iH_{\mathrm{eff}}^\dagger\,|\psi(t)\rangle
    =\kappa\langle\psi(t)|a^\dagger a|\psi(t)\rangle ,
    \label{eq:SM-norm}
\end{equation}
A jump from $|u_\mu,1\rangle$ completes the transition from $\nu$ to $\mu$. Once the quasi-stationary amplitude~\eqref{eq:SM-c3} is reached, the corresponding rate is
\begin{equation}
    \begin{aligned}
    \Gamma_{\mu,\nu}(\tau,t_0)
    &\equiv\kappa\,|c^{\mathrm{ss}}_{\mu,\nu}(\tau,t_0)|^2\\
    &\simeq g^2\kappa\sum_{n,m}
    \frac{M^{(n)}_{\mu,\nu}M^{(m)*}_{\mu,\nu}\,
          e^{i(n-m)\omega(t_0+\tau)}}
         {(\Delta E^{(n)}_{\mu,\nu}-i\kappa/2)
          (\Delta E^{(m)}_{\mu,\nu}+i\kappa/2)}.
    \end{aligned}
    \label{eq:SM-Gamma-t}
\end{equation}
The cross terms $n\neq m$ carry the phase $e^{i(n-m)\omega(t_0+\tau)}$ and average to zero when $t_0$ is integrated over one driving period. The physically observed rate is the average over one period of the section,
\begin{equation}
    \Gamma_{\mu,\nu}
    =\frac{1}{T}\int_0^T dt_0\,\Gamma_{\mu,\nu}(\tau,t_0),
\end{equation}
which is well defined once $\tau\gg\kappa^{-1}$, so that the quasi-stationary amplitude~\eqref{eq:SM-c3} has been reached. Using $\frac{1}{T}\int_0^Tdt_0\,e^{i(n-m)\omega t_0}=\delta_{nm}$, all cross terms between different channels vanish, and the stroboscopic average gives
\begin{equation}
    \Gamma_{\mu,\nu}
    = \sum_n \left|M^{(n)}_{\mu,\nu}\right|^2
    \frac{g^2\kappa}
         {(\varepsilon_\mu-\varepsilon_\nu+\Delta+n\omega)^2+(\kappa/2)^2}.
    \label{eq:SM-FGR}
\end{equation}
This is the Floquet Fermi golden rule, Eq.~\eqref{eq.FGR} of the main text. Its Lorentzian replaces the energy-conserving $\delta$-function of the closed-system rule~\eqref{eq:SM-FGRclosed}: the function $\kappa/(E^2+\kappa^2/4)$ tends to $2\pi\delta(E)$ as $\kappa\to0$. The rate is also independent of the choice of time origin: shifting the time origin only multiplies each Fourier coefficient by a phase, which cancels in $|M^{(n)}_{\mu,\nu}|^2$ and therefore leaves Eq.~\eqref{eq:SM-FGR} unchanged.

% Three limiting cases, used repeatedly in the main text, make the physical content explicit. (i) If a single channel $n_0$ is resonant, $\Delta E^{(n_0)}_{\mu,\nu}=0$, its Lorentzian dominates and $\Gamma_{\mu,\nu}\simeq(4g^2/\kappa)|M^{(n_0)}_{\mu,\nu}|^2$: a narrower cavity line yields a higher, sharper resonance. (ii) If no channel is resonant, every detuning is of order $\omega\gg\kappa$, and $\Gamma_{\mu,\nu}\sim(g^2\kappa/\omega^2)\sum_n|M^{(n)}_{\mu,\nu}|^2$, so a resonant channel is enhanced over a typical off-resonant one by $(\omega/\kappa)^2\gg1$. (iii) As $\kappa\to0$ the rate tends to zero: without a loss channel the photon is always reabsorbed and the spin never completes an irreversible transition, consistently with the $\delta$-function discussion above. These statements assume weak coupling $g$, the hierarchy $\omega\gg\kappa$, and a single resonantly dominant channel; near a quasienergy degeneracy, where two channels are simultaneously resonant, their competition must be analyzed separately, as done in the main text.

\subsection{Symmetry-protected selection rules}

The rate~\eqref{eq:SM-FGR} sums over infinitely many Fourier harmonics. We now show that a dynamical symmetry of the driven spin forces the odd Fourier components of the inter-state matrix elements to vanish, leaving only the transition channels with even driving photons~\cite{engelhardtDynamicalSymmetriesSymmetryProtected2021, Wang2021}. Consider the Floquet equation for a generic periodically driven Hamiltonian $H(t)=H(t+T)$,
\begin{equation}
    \left[H(t)-i\partial_t\right]|u_\mu(t)\rangle=\varepsilon_\mu|u_\mu(t)\rangle .
    \label{eq:SM-Floquet}
\end{equation}
A dynamical symmetry is an operator $\hat{\Sigma}$, together with parameters $(t_S,\beta_S,\alpha_S)$, that maps the Floquet operator onto itself up to a phase:
\begin{equation}
    \hat{\Sigma}\left[H(t_S+\beta_S t)-i\partial_t\right]\hat{\Sigma}^{-1}
    =\alpha_S\left[H(t)-i\partial_t\right].
    \label{eq:SM-symdef}
\end{equation}
When such a symmetry exists, it relates Floquet eigenstates at transformed times:
\begin{equation}
    |u_{\mu'}(t)\rangle=e^{i\varphi_\mu}\hat{\Sigma}|u_\mu(t_S+\beta_S t)\rangle,
    \label{eq:SM-symstate}
\end{equation}
with a phase $\varphi_\mu$ fixed by the symmetry.

Our model~\eqref{Fmodel} possesses a dynamical parity symmetry specified by
\begin{equation}
    (\hat{\Sigma},t_S,\beta_S,\alpha_S)=(\sigma_z,T/2,1,1).
\end{equation}
Indeed, shifting time by half a period flips the sign of the drive, $H_S(t+T/2)=A\cos(\omega t)\sigma_y+\frac{\Omega}{2}\sigma_z$, and conjugation by $\sigma_z$ flips it back because $\sigma_z\sigma_y\sigma_z=-\sigma_y$ while $\sigma_z\sigma_z\sigma_z=\sigma_z$:
\begin{equation}
    \sigma_z H_S(t+T/2)\sigma_z=H_S(t).
\end{equation}
Applied to the Floquet states, this symmetry implies the half-period relation
\begin{equation}
    \sigma_z|u_\mu(t+T/2)\rangle=e^{i\varphi_\mu}|u_\mu(t)\rangle,
    \label{eq:SM-halfperiod}
\end{equation}
where $\varphi_\mu=\pm\pi n_\mu$ with an integer $n_\mu$.

We use Eq.~\eqref{eq:SM-halfperiod} to evaluate the Fourier component~\eqref{eq:SM-Fourier}. Splitting the integral into the two halves of a period and changing variables $t\to t+T/2$ in the second half, we find
\begin{equation}
    \begin{aligned}
    M^{(n)}_{\mu,\nu}
    &=\frac{1}{T}\int_0^{T/2}dt\,M_{\mu,\nu}(t)e^{-in\omega t}
     +\frac{1}{T}\int_0^{T/2}dt\,M_{\mu,\nu}(t+T/2)e^{-in\omega(t+T/2)}.
    \end{aligned}
\end{equation}
Equation~\eqref{eq:SM-halfperiod}, together with $\sigma_z\sigma^-\sigma_z=-\sigma^-$, gives
\begin{equation}
    M_{\mu,\nu}(t+T/2)
    =e^{-i(\varphi_\mu-\varphi_\nu)}
     \langle u_\mu(t)|\sigma_z\sigma^-\sigma_z|u_\nu(t)\rangle
    =-e^{-i(\varphi_\mu-\varphi_\nu)}M_{\mu,\nu}(t),
\end{equation}
and since $e^{-in\omega T/2}=e^{-in\pi}=(-1)^n$, the two halves combine into
\begin{equation}
    M^{(n)}_{\mu,\nu}
    =\frac{1}{T}\left(1-e^{-i(\varphi_\mu-\varphi_\nu)}(-1)^n\right)
    \int_0^{T/2}dt\,M_{\mu,\nu}(t)e^{-in\omega t}.
    \label{eq:SM-Mselect}
\end{equation}
Equation~\eqref{eq:SM-Mselect} determines which Fourier harmonics are allowed once the relative symmetry phase $\varphi_\mu-\varphi_\nu$ is known. For the two-level system, $\sigma_y H_S^*(t)\sigma_y=-H_S(t)$ maps a Floquet state with quasienergy $\varepsilon_+$ to one with quasienergy $-\varepsilon_+$. Choosing the first Floquet zone such that $\varepsilon_-=-\varepsilon_+$, we can therefore write
\begin{equation}
    |u_-(t)\rangle=\eta\sigma_y|u_+(t)\rangle^*,\qquad |\eta|=1,
\end{equation}
where $\eta$ is an arbitrary time-independent phase reflecting the freedom to choose the phases of the two states. Using $\sigma_z\sigma_y=-\sigma_y\sigma_z$ and Eq.~\eqref{eq:SM-halfperiod}, we obtain
\begin{equation}
    \sigma_z|u_-(t+T/2)\rangle
    =-\eta\sigma_y\bigl[\sigma_z|u_+(t+T/2)\rangle\bigr]^*
    =-e^{-i\varphi_+}|u_-(t)\rangle.
\end{equation}
Thus $e^{i\varphi_-}=-e^{-i\varphi_+}$, independently of $\eta$. Since the half-period symmetry eigenvalues are $\pm1$, the two states have opposite eigenvalues, and their symmetry phases differ by $\pi$. For $\mu\ne\nu$, the prefactor in Eq.~\eqref{eq:SM-Mselect} is therefore $1+(-1)^n$, which vanishes for odd $n$. Consequently,
\begin{equation}
    M^{(n)}_{\mu,\nu}=0\quad\text{for odd }n,\qquad\mu\ne\nu,
    \label{eq:SM-even}
\end{equation}
and only terms with even $n$---transitions exchanging an even number of drive photons---contribute to the rate~\eqref{eq:SM-FGR}. This selection rule explains why the low-entropy stabilization branches in Fig.~\ref{contourf}(a) occur only at the even-$n$ resonance conditions of Eq.~\eqref{Resonance}.

We now show that every adjacent pair on the unfolded spin-$S$ Floquet ladder obeys the same even-harmonic selection rule. The key is to determine the half-period symmetry phases directly from the spin-$S$ evolution operator. Write $Z=R_z^{(S)}(\pi)=e^{-i\pi S_z}$ and denote the propagator of $H_S(t)=\vec B(t)\cdot\vec S$ by $\mathcal U(t_2,t_1)$. The symmetry $ZH_S(t+T/2)Z^\dagger=H_S(t)$ implies
\begin{equation}
    \mathcal U(T,T/2)=Z^\dagger\mathcal U(T/2,0)Z.
\end{equation}
Define $Q=Z\mathcal U(T/2,0)$, which combines evolution over half a period with the symmetry rotation. Since $Z^2=(-1)^{2S}\mathbb I$, multiplication of the two half-period propagators gives
\begin{equation}
    \mathcal U(T,0)=Z^\dagger\mathcal U(T/2,0)Z\mathcal U(T/2,0)
    =(-1)^{2S}Q^2.
    \label{eq:SM-Qsquare}
\end{equation}

Because the Hamiltonian is linear in $\vec S$, its propagator is a spin rotation, as is $Q$. Choose a rotation $R$ that aligns the rotation axis of $Q$ with the $z$ axis, so that
\begin{equation}
    Q=R e^{-i\theta S_z}R^\dagger.
\end{equation}
The factor $(-1)^{2S}\mathbb I$ in Eq.~\eqref{eq:SM-Qsquare} can be written as either $e^{i2\pi S_z}$ or $e^{-i2\pi S_z}$. We introduce $s=\pm1$ to retain this redundancy in the rotation-angle representation, writing $(-1)^{2S}\mathbb I=e^{i2\pi sS_z}$. Equation~\eqref{eq:SM-Qsquare} then becomes
\begin{equation}
    \mathcal U(T,0)=R e^{-i(2\theta-2\pi s)S_z}R^\dagger
    =R e^{-i\delta\varepsilon_F T S_z}R^\dagger,
    \qquad \theta=\frac{\delta\varepsilon_F T}{2}+s\pi.
\end{equation}
The two choices of $s$ give values of $(2\theta-2\pi s)/T$ that differ by $2\omega$ but represent the same one-period propagator. We choose the value corresponding to the unfolded spacing $\delta\varepsilon_F$ in Sec.~III.A. The associated periodic Floquet states are
\begin{equation}
    |u_{F,m}(t)\rangle=\mathcal U(t,0)R|m\rangle e^{im\delta\varepsilon_F t},
    \qquad m=-S,\ldots,S.
\end{equation}
The one-period propagator above ensures that these states are $T$-periodic and have quasienergies $m\delta\varepsilon_F$.

To evaluate their half-period symmetry phases, use
$Z\mathcal U(t+T/2,T/2)=\mathcal U(t,0)Z$, which follows from the same Hamiltonian symmetry. Composing with $\mathcal U(T/2,0)$ gives
\begin{equation}
    Z\mathcal U(t+T/2,0)=\mathcal U(t,0)Q.
\end{equation}
Substitution into the periodic states, followed by $QR|m\rangle=e^{-im\theta}R|m\rangle$, yields
\begin{equation}
    \begin{aligned}
    R_z^{(S)}(\pi)|u_{F,m}(t+T/2)\rangle
    &=\mathcal U(t,0)QR|m\rangle e^{im\delta\varepsilon_F(t+T/2)}\\
    &=e^{-im(\theta-\delta\varepsilon_F T/2)}|u_{F,m}(t)\rangle\\
    &=e^{-is\pi m}|u_{F,m}(t)\rangle.
    \end{aligned}
    \label{eq:SM-spinS-halfperiod}
\end{equation}
Thus neighboring unfolded states have opposite half-period symmetry eigenvalues.

To obtain the selection rule, define the transition matrix element
$M_{m',m}(t)=\langle u_{F,m'}(t)|S^-|u_{F,m}(t)\rangle$.
Solving Eq.~\eqref{eq:SM-spinS-halfperiod} for each state at $t+T/2$ and substituting into this matrix element gives
\begin{equation}
    \begin{aligned}
    M_{m',m}(t+T/2)
    &=e^{is\pi(m'-m)}
      \langle u_{F,m'}(t)|R_z^{(S)}(\pi)S^-R_z^{(S)\dagger}(\pi)|u_{F,m}(t)\rangle\\
    &=-e^{is\pi(m'-m)}M_{m',m}(t),
    \end{aligned}
\end{equation}
where the last step uses $R_z^{(S)}(\pi)S^-R_z^{(S)\dagger}(\pi)=-S^-$. For adjacent states, $m'-m=\pm1$, so $e^{is\pi(m'-m)}=-1$ and the two signs cancel: $M_{m',m}(t+T/2)=M_{m',m}(t)$. Splitting its Fourier integral into two half-periods, as in Eq.~\eqref{eq:SM-Mselect}, now yields
\begin{equation}
    M_{m',m}^{(n)}
    =\frac{1+(-1)^n}{T}\int_0^{T/2}dt\,M_{m',m}(t)e^{-in\omega t},
    \qquad m'-m=\pm1.
\end{equation}
Hence odd harmonics vanish on every adjacent link of the unfolded ladder. Since all these links have quasienergy differences $\pm\delta\varepsilon_F$, their sideband-resonance conditions are
\begin{equation}
    \Delta=\pm\delta\varepsilon_F+q\omega,\qquad q=-n\in2\mathbb Z,
\end{equation}
including the Red resonance condition $\Delta=-\delta\varepsilon_F+q\omega$.

Folding a state according to $\varepsilon_\mu=\varepsilon_{F,m}-\ell_m\omega$ multiplies its periodic part by $e^{-i\ell_m\omega t}$ and shifts its half-period symmetry phase by $\pi\ell_m$ modulo $2\pi$. An odd difference between the zone shifts of two neighboring states therefore selects odd Fourier indices in the folded basis. Thus the integer in $\Delta=-\delta\varepsilon_{\mu,\nu}+q'\omega$ need not be even, although the common unfolded condition above retains even $q$.

\section{Robustness to dephasing}

So far, spin relaxation has been the only decoherence channel acting on the spin. In any experimental realization, however, dephasing is unavoidable, and it may destroy the phase coherence required for the preparation of Floquet eigenstates. In this section we show numerically that cavity-assisted stabilization survives moderate dephasing, and we compare dephasing and spin relaxation as the two rates are varied.

\begin{figure}[h]
    \centering
    \includegraphics[width=0.6\linewidth]{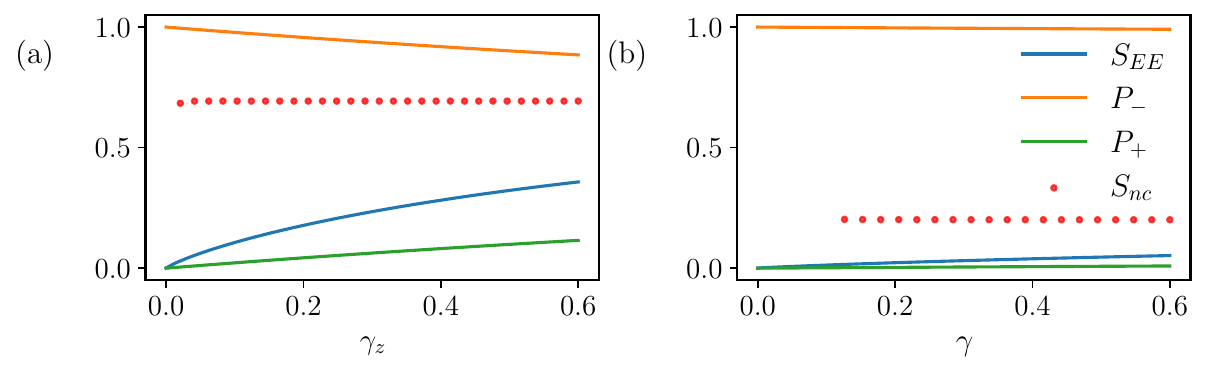}
    \caption{(a) Entropy and fidelity of the periodically driven two-level system coupled to the cavity, as functions of the dephasing rate $\gamma_z$. (b) Entropy and fidelity as functions of the spin-relaxation rate $\gamma$. The red dots show $S_{nc}$, the entropy of the spin with the cavity removed. Here we choose the parameters $A=60$, $\Omega=100$, and $g=\kappa=0.05A$. } 
    \label{gammaz}
\end{figure}

We use the two-level FTS model of Sec.~II and add a dephasing channel to the master equation~\eqref{rhoME},
\begin{equation}
    \gamma_z\,\mathcal D_{\sigma^z}[\rho]
    =\gamma_z\bigl(\sigma^z\rho\sigma^z-\rho\bigr),
\end{equation}
where $\gamma_z$ is the dephasing rate. This Lindblad term randomizes the relative phase of the two spin states without changing their populations. We compute the long-time von Neumann entropy and the fidelity with the target Floquet eigenstate as functions of $\gamma_z$ in Fig.~\ref{gammaz}(a). For comparison, Fig.~\ref{gammaz}(b) shows the same observables under variation of the spin-relaxation rate $\gamma$. For weak noise, the entropy remains low and the fidelity takes a high value. Therefore, the dissipative cavity continues to stabilize the spin, and stabilization is lost when the noise rate $\gamma_z,\gamma$ becomes comparable to the coupling and loss rates $g$ and $\kappa$.

To isolate the role of the cavity, the red dots in Fig.~\ref{gammaz} show the asymptotic entropy when the cavity is removed. Without it, dephasing drives the spin toward the maximally mixed state, $S=\ln 2$, whereas spin relaxation produces a high-entropy state that is nevertheless not fully mixed; with the cavity, by contrast, the entropy stays low for weak noise in both panels. The control confirms that the low-entropy state is actively maintained by the cavity.

\section{Why the competing transition is absent in the rotating-field model}
\label{sec:SM3}

The main text states that the Red--Blue competition is general but may be absent on the specific drive. Here we show analytically why it is absent in the exactly solvable rotating-field model of Ref.~\cite{ritterAutonomousStabilizationFloquet2025}: the Fourier spectrum of its transition matrix elements is restricted to three sidebands, and none of the Red sidebands can satisfy the resonance condition at a positive cavity frequency.

The spin is driven by a circularly polarized field in the $xz$ plane,
\begin{equation}
    H_B(t)=\frac{B_0}{2}\bigl[\cos(\omega t)\,\sigma_z+\sin(\omega t)\,\sigma_x\bigr],
    \label{eq:SM3-HB}
\end{equation}
and is statically coupled to a lossy cavity as in Eq.~\eqref{Fmodel}. Passing to the rotating frame $|\psi_{\rm rot}(t)\rangle=U(t)|\psi(t)\rangle$, with $U(t)=e^{i\omega t\sigma_y/2}$, renders the spin Hamiltonian static,
\begin{equation}
    H_{\rm rot}=U H_B U^\dagger+i\dot U U^\dagger
    =\frac12(B_0\sigma_z-\omega\sigma_y)
    =\frac{\Omega_R}{2}\bigl(\cos\vartheta\,\sigma_z-\sin\vartheta\,\sigma_y\bigr),
    \label{eq:SM3-Hrot}
\end{equation}
where
\begin{equation}
    \Omega_R=\sqrt{B_0^2+\omega^2},\qquad
    \cos\vartheta=\frac{B_0}{\Omega_R}\equiv\beta,\qquad
    \sin\vartheta=\frac{\omega}{\Omega_R}\equiv\alpha .
\end{equation}
The Floquet states are $|\phi_\pm(t)\rangle=e^{i\omega t/2}U^\dagger(t)|\chi_\pm\rangle$, where $|\chi_\pm\rangle$ are the eigenstates of $H_{\rm rot}$. The phase $e^{i\omega t/2}$ compensates for $U(t+T)=-U(t)$, giving quasienergies $\varepsilon_\pm=\omega/2\pm\Omega_R/2$ and hence $\delta\varepsilon=\Omega_R$ before folding. This common phase cancels in the transition matrix elements.

To evaluate the transition amplitudes, transform $\sigma^-$ together with the Floquet states into the rotating frame. Because a rotation about $y$ mixes $\sigma^-$ with $\sigma^+$ and $\sigma_z$,
\begin{equation}
    U\sigma^-U^\dagger
    =\frac{1+\cos\omega t}{2}\,\sigma^-
    -\frac{1-\cos\omega t}{2}\,\sigma^+
    +\frac{\sin\omega t}{2}\,\sigma_z .
\end{equation}
Taking the matrix elements between the tilted spinors, with phases chosen as $|\chi_+\rangle=i e^{-i\vartheta\sigma_x/2}|\uparrow\rangle$ and $|\chi_-\rangle=e^{-i\vartheta\sigma_x/2}|\downarrow\rangle$, then gives
\begin{equation}
\begin{aligned}
    M_b(t)&=\langle\phi_-(t)|\sigma^-|\phi_+(t)\rangle
    =\frac{i}{2}\bigl(\beta+\cos\omega t\bigr)-\frac{\alpha}{2}\sin\omega t,\\
    M_r(t)&=\langle\phi_+(t)|\sigma^-|\phi_-(t)\rangle
    =\frac{i}{2}\bigl(\beta-\cos\omega t\bigr)-\frac{\alpha}{2}\sin\omega t .
\end{aligned}
    \label{eq:SM3-Mt}
\end{equation}
Each amplitude is a constant plus the first harmonic alone. Its Fourier spectrum therefore contains only $n=0,\pm1$, with carrier weight $|M_{b,r}^{(0)}|^2=\beta^2/4$, while every sideband with $|n|\ge2$ vanishes identically---at any driving frequency. This is the kinematic selection rule of circular polarization: the micromotion is a uniform rigid rotation and contains no higher harmonics that a multiphoton transition could borrow. (We note that the period-averaged total weights of the two amplitudes are equal; the asymmetry found below is therefore not a matter of total dipole strength but of the sideband structure.)

It remains to combine this spectrum with the zero-detuning condition in Eq.~\eqref{eq:SM-FGR}, written as $\Delta=\pm\Omega_R+q\omega$ with $q=-n$. Here $n$ is the Fourier index, and the allowed resonance labels are $q=0,\pm1$. For the Blue process, all three allowed sidebands place the resonance at a positive cavity frequency,
\begin{equation}
    \Delta_b^{(0)}=\Omega_R,\qquad
    \Delta_b^{(-1)}=\Omega_R-\omega>0,\qquad
    \Delta_b^{(+1)}=\Omega_R+\omega ,
\end{equation}
and the carrier resonance $\Delta=\Omega_R$ is carried by the strong $n=0$ component. For the Red process, in contrast,
\begin{equation}
    \Delta_r^{(0)}=-\Omega_R<0,\qquad
    \Delta_r^{(+1)}=\omega-\Omega_R<0,\qquad
    \Delta_r^{(-1)}=-\Omega_R-\omega<0 .
\end{equation}
The two-drive-photon Red resonance would be $\Delta_r^{(+2)}=2\omega-\Omega_R$, which requires absorbing two drive photons and exists only for $\omega>B_0/\sqrt3$; but its $n=-2$ Fourier component, like every $|n|\ge2$ component, is forbidden by Eq.~\eqref{eq:SM3-Mt}. No Red resonance therefore exists at a positive cavity frequency at any drive frequency.

The remaining off-resonant contribution is negligible. At the Blue carrier resonance $\Delta=\Omega_R$, the Floquet FGR rate of Eq.~\eqref{eq:SM-FGR} gives
\begin{equation}
    \Gamma_b\simeq\frac{g^2\kappa\,|M_b^{(0)}|^2}{(\kappa/2)^2}
    =\frac{g^2\beta^2}{\kappa},
\end{equation}
whereas the Red detunings are $2\Omega_R+n\omega>\Omega_R$ for $n=0,\pm1$, so that
\begin{equation}
    \frac{\Gamma_r}{\Gamma_b}\sim\left(\frac{\kappa}{4\beta\Omega_R}\right)^2\ll1
\end{equation}
in the regime $\kappa\ll B_0\lesssim\Omega_R$ of Ref.~\cite{ritterAutonomousStabilizationFloquet2025}. The Blue process thus operates alone throughout the finite-frequency range studied there.

For a generic periodically driven spin, the micromotion need not be a uniform rotation. In our linearly polarized model, the micromotion contains all harmonics; the dynamical parity symmetry of SM Sec.~1 suppresses only the odd-$n$ sidebands, while the $n=\pm2,\pm4$ components remain finite. At high driving frequencies these components carry negligible weight and the resonance condition alone selects the stabilized state, as in the rotating-field model. At intermediate frequencies, however, the $n=-2$ Fourier component acquires substantial weight and opens the Red resonance $\Delta=-\delta\varepsilon+2\omega$ at a positive cavity frequency; near $\delta\varepsilon\simeq\omega$ the Red and Blue resonances even approach each other. The competing process is then unavoidable, and stabilization succeeds only when the resonant channel and the dominant matrix element coincide, as discussed in the main text.

\end{document}